\documentclass[11pt]{article}

\newcommand{\nc}{\newcommand}

\nc{\eqr}[1]{(\ref{#1})}
\nc{\sref}[1]{\S~\ref{#1}}
\nc{\tref}[1]{Table~\ref{#1}}
\nc{\fref}[1]{Figure~\ref{#1}}
\nc{\cref}[1]{Chapter~\ref{#1}}
\nc{\beq}{\begin{equation}}
\nc{\eeq}{\end{equation}}
\nc{\barray}{\begin{eqnarray}}
\nc{\earray}{\end{eqnarray}}
\nc{\barrayn}{\begin{eqnarray*}}
\nc{\earrayn}{\end{eqnarray*}}
\nc{\bcenter}{\begin{center}}
\nc{\ecenter}{\end{center}}
\nc{\onehalf}{\frac{1}{2}}
\nc{\lra}{\longrightarrow}
\nc{\ra}{\rightarrow}
\nc{\mod}{\mbox{mod }}
\nc{\pf}{{\sc Proof: }}
\nc{\setall}{\setcounter{equation}{0}
        \setcounter{definition}{0}
        \setcounter{lemma}{0}
        \setcounter{convention}{0}
        \setcounter{conjecture}{0}
        \setcounter{theorem}{0}
        \setcounter{assertion}{0}
        \setcounter{digression}{0}
        \setcounter{property}{0}
        \setcounter{fact}{0}
        \setcounter{corollary}{0}}
\nc{\setequation}{\setcounter{equation}{0}}

\def\sla#1{\raise.15ex\hbox{/}\kern-.57em #1}
\def\slas#1{\raise.15ex\hbox{/}\kern-.62em #1}
\nc{\tbyt}[4]{\left( \begin{array}{rr}
        #1 & #2 \\
        #3 & #4
        \end{array}\right)}
\nc{\abcd}{\left( \begin{array}{cc}
        a & b \\
        c & d
        \end{array}\right)}
\nc{\inner}[2]{\langle #1 , #2 \rangle}
\nc{\e}[1]{{\mbox e}^{#1}}
\nc{\met}[2]{g_{#1 #2}}
\nc{\oover}[1]{\frac{1}{#1}}
\nc{\wed}[2]{ #1 \wedge #2}
\nc{\bhat}[1]{\hat{\mbox{\boldmath $#1$}}}
\nc{\mbold}[1]{\mbox{\boldmath $#1$}}

\def\sCC{{\kern 0.27em\vrule height1.45ex width0.03em depth0em
	  \kern-0.30em\rm C}}
\def\C{{\mathchoice
  {\sCC}
  {\sCC}
  {\kern 0.225em \vrule height1.05ex width0.025em depth0em \kern-0.25em \rm C}
  {\kern 0.180em \vrule height0.78ex width0.02em depth0em \kern-0.2em \rm C}
	}}
\def\sHH{{\rm I\kern-.16em{}H}}
\def\H{{\mathchoice
  {\sHH}
  {\sHH}
  {\rm I\kern-.13em{}H}
  {\rm I\kern-.13em{}H} }}
\def\sNN{{\rm I\kern-.16em{}N}}
\def\N{{\mathchoice
  {\sNN}
  {\sNN}
  {\rm I\kern-.12em{}N}
  {\rm I\kern-.10em{}N} }}
\def\sPP{{\rm I\kern-.16em{}P}}
\def\P{{\mathchoice
  {\sPP}
  {\sPP}
  {\rm I\kern-.12em{}P}
  {\rm I\kern-.10em{}P} }}
\def\sQQ{{\kern 0.27em \vrule height1.45ex width0.03em depth0em
	  \kern-0.30em \rm Q}}
\def\Q{{\mathchoice
	{\sQQ}
	{\sQQ}
  {\kern 0.225em \vrule height1.05ex width0.025em depth0em \kern-0.25em \rm Q}
  {\kern 0.180em \vrule height0.78ex width0.020em depth0em \kern-0.20em \rm Q}
	}}
\def\sRR{{\rm I\kern-0.16em{}R}}
\def\R{{\mathchoice
  {\sRR}
  {\sRR}
  {\rm I\kern-0.12em{}R}
  {\rm I\kern-0.10em{}R} }}
\def\sZZ{{\rm Z\kern-0.32em{}Z}}
\def\Z{{\mathchoice
  {\sZZ}
  {\sZZ} 
  {\rm Z\kern-0.3em{}Z}     %.3
  {\rm Z\kern-0.25em{}Z} }}  %.25
\def\ZZZ{{\rm Z\kern-0.24em{}Z}}

\nc{\vE}{\vec{E}}
\nc{\vB}{\vec{B}}

\nc{\cA}{{\cal A}}
\nc{\cB}{{\cal B}}
\nc{\cC}{{\cal C}}
\nc{\cD}{{\cal D}}
\nc{\cE}{{\cal E}}
\nc{\cF}{{\cal F}}
\nc{\cG}{{\cal G}}
\nc{\cH}{{\cal H}}
\nc{\cI}{{\cal I}}
\nc{\cJ}{{\cal J}}
\nc{\cK}{{\cal K}}
\nc{\cL}{{\cal L}}
\nc{\cM}{{\cal M}}
\nc{\cN}{{\cal N}}
\nc{\cO}{{\cal O}}
\nc{\cP}{{\cal P}}
\nc{\cQ}{{\cal Q}}
\nc{\cR}{{\cal R}}
\nc{\cS}{{\cal S}}
\nc{\cU}{{\cal U}}
\nc{\cV}{{\cal V}}
\nc{\cW}{{\cal W}}
\nc{\cX}{{\cal X}}
\nc{\cY}{{\cal Y}}
\nc{\cZ}{{\cal Z}}

\nc{\bA}{{\bf A}}
\nc{\bB}{{\bf B}}
\nc{\bE}{{\bf E}}
\nc{\bI}{{\bf I}}
\nc{\bJ}{{\bf J}}
\nc{\bK}{{\bf K}}
\nc{\bR}{{\bf R}}
\nc{\bZ}{{\bf Z}}

\nc{\al}{\alpha}
\nc{\be}{\beta}
\nc{\ga}{\gamma}
\nc{\de}{\delta}
\nc{\ep}{\epsilon}
\nc{\n}{\nu}
\nc{\m}{\mu}
\nc{\sskip}{\vspace{5mm}}
\nc{\nskip}{\vspace{-2mm}}
\nc{\vs}[1]{\vspace{#1}}
\nc{\hs}{\hspace{1cm}}
\nc{\hshalf}{\hspace{5mm}}

\ifx\undefined\square
  \def\square{\vrule width.6em height.5em depth.1em\relax}\fi
\def\qed{\ifhmode\unskip\nobreak\fi\quad
  \ifmmode\square\else$\m@th\square$\fi}
\newtheorem{claim}{\large\bf CLAIM}

\nc{\ds}[1]{\tau_{#1}}
\nc{\cc}[4]{C^{#1}_{#2}(#3,#4)}
\nc{\dd}[4]{D^{#1}_{#2}(#3,#4)}
\nc{\ccor}[2]{{\langle #1 \rangle_{\mbox{\tiny $#2$}}}}
\nc{\cor}[2]{{\langle\hspace{-1mm}\langle #1 \rangle\hspace{-1mm}\rangle}_{#2}}
\nc{\ms}[2]{{\cal M}_{#1}(#2)}
\nc{\sig}[1]{\sigma_{#1}}

\renewcommand{\thefootnote}{\fnsymbol{footnote}}
\begin{document}

\begin{titlepage}
{\flushright{\small MIT-CTP-2932\\hep-th/9912078\\}}
\begin{center}
\vspace{5mm}

{\LARGE Descendant Gromov-Witten Invariants,\\ \vspace{3mm}
 Simple Hurwitz Numbers, and the Virasoro Conjecture for $\P^1$}\\
\vspace{3mm}
\end{center}

\vspace{.5cm}
\begin{center}

{\large\sc Jun S. Song\footnote{E-mail:
jssong@mit.edu.  Research supported in part
by the NSF Graduate Fellowship and the U.S. Department of Energy under cooperative research
agreement $\#$DE-FC02-94ER40818.
}}\\
\vspace{2mm}
{\it Center for Theoretical Physics}\\
{\it Massachusetts Institute of Technology}\\
{\it Cambridge, Massachusetts 02139, USA}
\end{center}

\vspace{1cm}
\begin{abstract}
In this ``experimental'' research, we use
known topological recursion relations in genera-zero, -one, and
-two to compute the $n$-point descendant Gromov-Witten invariants of 
$\P^1$ for arbitrary degrees and low values of $n$.  The results are
consistent with the Virasoro conjecture and also lead
to explicit computations of all Hodge integrals in these genera.  We also
derive new recursion 
relations for simple Hurwitz numbers similar to those of Graber and
Pandharipande.
\end{abstract}
\end{titlepage}

\renewcommand{\thefootnote}{\arabic{footnote}}

%%%%%%%%%%%%%########################################
%  Introduction 
%%%%%%%%%%%%%########################################

\section{Introduction}
It is well-known 
that the intersection theory on the compactified moduli space
$\overline{\cM}_{g,n}$ 
of stable $n$-pointed
genus-$g$ curves, or equivalently the two-dimensional pure gravity, is
governed by an integrable KdV 
hierarchy \cite{K,Witten}.   More precisely, the KdV hierarchy allows
one to compute recursively 
the intersection numbers of tautological divisors on
$\overline{\cM}_{g,n}$, for 
arbitrary $g$ and $n$,
  in terms of two basic invariants in genus-0 and
genus-1.  Physically, it means the following:  It is a common and
useful practice to perturb a given quantum field theory by introducing
into the action couplings to physical operators and to study the perturbed
partition function which becomes the generating function for the
correlators of the original theory.  For example, in
a topological string theory on a target space $V$, the physical
operators are the cohomology classes $\gamma_a$ of $V$ and their gravitational
descendants $\ds{m,a}, m\in\Z_{\geq 1}$.  In this case, one considers
a perturbation by 
	\beq
	\sum^{\infty}_{m=0}\sum_{\gamma_a\in H^*(V)}t^a_m\,\ds{m,a}
	\eeq
where $\ds{0,a}$ represents the {\it primary field} associated with the
cohomology class $\gamma_a$ itself.  The parameters $t^a_m$ are said to form the
coordinates on the so-called {\it large phase
space}\footnote{Similarly, the {\it small phase space}\/ refers to a
space of deformations by only the primary fields; i.e. the subspace
$t^a_m =0, m>0$, of
the large phase space.}.  In this setting,
the KdV structure  implies that the partition function of the perturbed
topological string theory on a point target space, in which the
puncture or identity operator is the only primary field,  
is a $\tau$-function of the hierarchy and the $\tau$-function is
uniquely fixed by the string equation.

 It was soon realized that the 
statement of the integrable structure can be rephrased in
terms of certain differential operators on the large phase space
which annihilate the $\tau$-function.  It turns out that these
operators furnish a representation of a subsector of the Virasoro
algebra, and thus the KdV hierarchy of the intersection theory
 is also known as the Virasoro constraints \cite{Dijkgraaf}.
 
One immediate generalization of the above picture is to introduce more
primary fields by coupling the  two-dimensional topological 
gravity to topological
field theories.  For example, coupling the topological minimal models
to topological gravity leads to $d<1$ topological string theories
which are governed by $W$-algebra constraints, generalizing the
Virasoro algebra.   A more interesting and perhaps more physical
 way is to consider topological
string theories  on more general non-trivial target spaces.  
This approach has led to  physical means of studying
the Gromov-Witten (GW) invariants, of which the subject of quantum
cohomology is a subset, on Fano and Calabi-Yau manifolds.  Such
physical models describe the intersection theory, sometimes called the
gravitational quantum cohomology, on some suitably
defined moduli spaces of stable holomorphic maps from Riemann surfaces
to target spaces \cite{Behrend,BM,KM,LT}.  Based on the previous
examples, should one
expect some kind of an integrable structure to govern the intersection theory in
these cases as well?

Through a
series of papers \cite{EY,EKY,EHX,EHX2,EX,Hori}, it has been indeed
conjectured  that
there should exist a certain integrable hierarchy which underlies the
gravitational quantum cohomology and which manifests itself again in
terms of a set of differential operators 
forming a half branch of  the Virasoro algebra:
	\beq
	\left[ L_n, L_m\right] = (n-m) L_{n+m},\ \ n,m\geq -1 \ .
	\eeq  
This conjecture is now referred to as the Virasoro conjecture and has
been proven up to 
genus-1 by mathematicians for manifolds satisfying certain
conditions \cite{DZ,Tian,Liu}; in particular, for complex projective
spaces  $\P^n$.
Historically, this conjecture is based on the
discovery of a matrix model for the topological string theory on $\P^1$
\cite{EY,EKY,Hori}; the Ward identities for the matrix model form a
Virasoro algebra.  The authors of \cite{EKY} have checked for a few
cases that the intersection numbers on the moduli space of stable maps
indeed satisfy the constraints implied by their conjecture.  Despite
some curious matchings, there is yet no complete proof\footnote{That
is, except
for a point and Calabi-Yau varieties of dimension greater than or
equal to three.} of their
conjecture even for $\P^1$.

In this paper,  we take a retrograding step towards attempting to unravel the
mystery of the Virasoro constraints for $\P^1$.
At first sight,
this example appears to be the simplest generalization of the pure
gravity case, but it turns out that the only interesting GW invariants
for $\P^1$ are the gravitational descendants.  The reason is that the
only primary fields of the theory are the identity and the second
cohomology class which can be eliminated via the puncture and divisor
equations \cite{Hori,Witten}.
We thus cannot obtain any nontrivial
information by restricting our attention to the small phase space, as
was done in \cite{EHX2,EX}, and
we would need to consider various descendant GW-invariants to study
whether there exist any possible constraints on the theory.  
Important ingredients in our computations of the GW-invariants are the
known topological recursion relations (TRRs) in genera-zero, -one and
-two.

Incidentally, the relation between the TRRs and the Virasoro
constraints are not clear, even in the present case of $\P^1$.  In the
pure gravity case, the Virasoro constraints, or the KdV hierarchy,
completely determine all the correlators in all genera in terms of
$\ccor{\ds{0,0} \ds{0,0}\ds{0,0}}{0}$ and $\ccor{\ds{1,0}}{1}$, and
there is no need for additional TRRs; the TRRs are thus redundant for
a point target space.  For higher dimensional target spaces, however,
the Virasoro constraints by themselves are not powerful enough to
determine all the correlators and require the help of additional
TRRs.  In fact, it is not known whether the Virasoro constraints, even
if they were true, together with various TRRs, 
would be able to determine all the correlators for a non-trivial
target space.  Interestingly, for $\P^1$, the positive modes of the 
Virasoro constraints are not needed to compute all the descendant
GW-invariants up to genus-2.  That is, the TRRs of 
\cite{EHX,Getzler2,Witten}, 
 together with the $L_{-1}$ and $L_{0}$ constraints\footnote{These two
constraints are proven to hold for all manifolds.  The $L_{-1}$
constraint is the string equation of Witten \cite{Witten}, and
$L_{0}$ the equation of Hori \cite{Hori} combining the dilaton,
divisor and dimension equations.}, are enough to
compute all the correlators in genus-zero, -one, and -two.  At least
in these low genera, it thus seems that the Virasoro constraints for
$\P^1$ are redundant, and indeed, we have checked for many of the
GW-invariants which
we have obtained via TRRs that they actually do satisfy the constraints.

Before we proceed, it is perhaps necessary to clarify the nature of
our work, so as to align the reader's line of thinking  with our
own.  
The philosophy of this paper is not to prove any parts of the Virasoro
conjecture.  Instead, we admittedly take an  un-innovative
approach to computing the descendant GW-invariants of $\P^1$ by using
the available 
topological recursion relations, and the numbers that we thus obtain are
independent of the Virasoro conjecture.  Since a genus-$g$ recursion
relation involves lower genus contributions, a mistake in genera-zero
and -one would propagate through any
subsequent computations in higher genera.  We therefore check
many\footnote{A computer program has allowed us to check that  over
10,000 Virasoro constraints are satisfied.} of 
our results by verifying that they satisfy the Virasoro constraints in
genera-zero and -one, 
which are rigorously proven to hold \cite{DZ,Tian,Liu}.  Based on
those numbers, we are able to compute the genus-2 
GW-invariants containing up to three arbitrary descendant fields, and we
again check that they satisfy many of the genus-2 Virasoro
constraints.  As the Virasoro constraints in genus-2 are conjectural,
our verification provides a minor support for the claim.

This paper is organized as follows:  We first compute the descendant
GW-invariants and the Hodge integrals in genera-zero, -one, and -two
just by using known topological recursion relations.   In
\S\ref{sec:Virasoro}, we use these results to check the Virasoro
conjecture by explicitly checking that the correlators satisfy the 
constraints.  We also comment on the higher-genus cases and on the
TRRs of Eguchi and Xiong \cite{EX}.  In \S\ref{sec:Hurwitz}, which is
independent of other sections, we derive
new recursion relations for simple Hurwitz numbers in genera-zero and
-one by using TRRs as well as  by applying the Virasoro constraints
discussed in the previous sections.  The paper concludes with
speculations and open questions regarding the relation between the
TRRs and the Virasoro constraints.

\vspace{1cm}
\noindent
\underline{NOTATIONS}

Many different notations are being used by mathematicians and
physicists.  Here, we clarify the conventions that we use:

\vspace{5mm}
\begin{tabular}{lll}
$\ds{m,\alpha}$ &\hspace{5mm} &the $m$-th descendant of a primary field 
		$\gamma_\alpha\in H^{2\alpha}(\P^1,\C)$.  See also
		\eqr{eq:GW-integral}.\\
$\ds{0,0}$ &\hspace{5mm}  & the identity element in $H^*(\P^1,\C)$.\\
$\ds{0,1}$&\hspace{5mm}  &  the basis of $H^2 (\P^1,\Z)$.\\
$t^{\alpha}_m$ & \hspace{5mm} & the coordinate associated with $\ds{m,\alpha}$
		on the large phase space. \\
$\cor{\mbox{ }}{g}$ & \hspace{5mm}& a genus-$g$ correlator in the large
		phase space.\\
$\ccor{\mbox{ }}{g}$ & \hspace{5mm}& a genus-$g$ correlator at the
		origin of  the large phase space, i.e. $t^a_m =0,
		\forall a,m$.\\
$\ccor{\mbox{ }}{g,d}$ & \hspace{5mm}& a degree-$d$ Gromov-Witten
		invariant in genus-$g$ at  $t^a_m =0,
		\forall a,m$.\\
$c_m$ & \hspace{5mm}& $\sum^m_{k=1} 1/k$.
\end{tabular}

\vspace{5mm}
\noindent
\begin{itemize}
	\item We call $m$ the {\it degree}\footnote{This degree should not be
confused with the degree of a stable map in
$\overline{\cM}_{g,n}(\P^1,d)$.} of the descendant $\ds{m,\alpha}$.
	\item Following the physics nomenclature, we sometimes call
	the $n$-point descendant invariant
	$\ccor{\ds{m_1,\alpha_1 }\cdots\ds{m_n,\alpha_n } }{g}$ an
	$n$-point correlation function, or simply an $n$-point
	functions. 
	\item Technically, an $n$-point GW-invariant
$\ccor{\ds{m_1,\alpha_1 }\cdots\ds{m_n,\alpha_n } }{g}$ is a sum of
Gromov-Witten invariants $\ccor{\ds{m_1,\alpha_1
}\cdots\ds{m_n,\alpha_n } }{g,d}$ in various degrees with coefficients in the
Novikov ring of $\P^1$, but since each correlator receives a contribution from
only one specific  degree, we will often use 
$\ccor{\ds{n_1,\alpha_1 }\cdots\ds{n_k,\alpha_k } }{g}$ and
$\ccor{\ds{n_1,\alpha_1 }\cdots\ds{n_k,\alpha_k } }{g,d}$
interchangeably.  The degree $d$ of the non-vanishing GW-invariant 
can be worked out easily from the
dimension of the virtual fundamental class
$[\overline{\cM}_{g,n}(\P^1,d)]^{\mbox{vir}}$.
\item Many terminologies from algebraic geometry are used in this
paper without explanation.  We will pretend that we know what they mean
and refer the reader to the available references for their definitions
\cite{CK,Getzler,GH}.
\end{itemize}

%%%%%%%%%%%%%%%%%#########################################
%   Properties of GW-invariants
%%%%%%%%%%%%%%%%%#########################################
\setequation
\section{Computations of the Descendant GW-Invariants}\label{sec:computation}

In this section, we compute the descendant GW-invariants of $\P^1$ in
genera-zero, -one and -two by
using the topological recursion relations of Witten \cite{Witten},
Eguchi-Hori-Xiong \cite{EHX} and
Getzler \cite{Getzler2}.
In general, a TRR in genus-$g$
involves lower genus GW-invariants; as a result, the computational 
usefulness of a TRR  depends on the knowledge of the lower
genus invariants.    We thus proceed systematically 
from descendant GW-invariants in
genus-0 to those in genus-2.

\subsection{Properties of the Descendant GW-Invariants}\label{sec:properties}
The descendant GW-invariants, also known as 
the gravitational correlators,
satisfy certain topological axioms which will be used throughout this
paper.   In this subsection, we briefly review these properties and
refer the reader to \cite{CK} for details.

Let $\overline{\cM}_{g,n}(V,\beta)$ be the compactified moduli space of
stable holomorphic maps $f:\Sigma\ra V$ of genus-$g$ $n$-pointed
Riemann surfaces  $\Sigma$ to a smooth projective variety $V$, such
that $f_* [\Sigma] = \beta \in H_2 (V,\Z)$.  Let
$\pi: \overline{\cM}_{g,n+1}(V,\beta) \ra \overline{\cM}_{g,n}(V,\beta)$
be the universal curve with $n$ natural sections 
	\beq
	\sigma_i : \overline{\cM}_{g,n}(V,\beta) \lra
\overline{\cM}_{g,n+1}(V,\beta) 
	\eeq
associated with the $n$ marked points.  The tautological line bundle
$\cL_i\ra \ \overline{\cM}_{g,n}(V,\beta)$ is defined to be
$\sigma^*_i \omega$, where  $\omega$ is the relative dualizing sheaf
of $\pi$, and we denote its first Chern class, called the tautological
$\psi$-class, by $\psi_i$.
	
	Let $\mbox{ev}: \overline{\cM}_{g,n}(V,\beta)
	\ra V^n$ be the evaluation map defined by
	\beq
	\mbox{ev}: [f:(\Sigma,z_1,\ldots,z_n)\ra V] \in
	\overline{\cM}_{g,n}(V,\beta)
 \longmapsto
	(f(z_1) , \ldots, f(z_n)) \in V^n. 
	\eeq
Then, the descendant GW-invariant is defined to be
	\beq \label{eq:GW-integral}
	\ccor{\ds{m_1,\alpha_1} \cdots \ds{m_n,\alpha_n}}{g,\beta} := 
	\int_{  [\overline{\cM}_{g,n}(V,\beta)]^{\mbox{\tiny vir}}}
	\psi^{m_1}_1 \cdots \psi^{m_n}_n \cup \mbox{ev}^*
	( \gamma_{\alpha_1} \otimes \cdots \otimes \gamma_{\alpha_n}) .
	\eeq
These invariants satisfy  certain ``axioms'' which are
generalizations of those occurring in the pure gravity case, i.e. in
the case of a point target space.  Specializing to the case of $\P^1$,
	they are:
\begin{itemize}
\item  {\bf Degree Axiom.}  The GW-invariant $\ccor{\ds{m_1,\alpha_1}
\cdots \ds{m_n,\alpha_n}}{g,d}$ vanishes if
\beq
	2d + 2g -2 + n \neq  \sum^n_i (m_i + \alpha_i ),
\eeq
	where $d\in\Z_{\geq 0}$ is the degree of the map,
i.e. $f_*[\Sigma] 
= d\, \beta$ where $\beta$ generates the effective cycles of
$H_2(\P^1,\Z)$. 
\item {\bf String Axiom.} For either $n+2g \geq 3$ or $d>1,n\geq
1$,  
	\beq
	\ccor{\ds{0,0}\ds{m_1,\alpha_1} \cdots
\ds{m_n,\alpha_n}}{g,d} = \sum^n_{i=1} \ccor{\ds{m_1,\alpha_1} 
\cdots\ds{m_{i-1},\alpha_{i-1}}
\ds{m_{i}-1,\alpha_i}\ds{m_{i+1},\alpha_{i+1}}
 \cdots\ds{m_n,\alpha_n}}{g,d}.
	\eeq
\item  {\bf Divisor Axiom.} For either $n+2g \geq 3$ or $d>1,n\geq
1$, 
	\barray
	\ccor{\ds{0,1}\ds{m_1,\alpha_1} \cdots
\ds{m_n,\alpha_n}}{g,d} &=&  
\sum_i^{n}
\ccor{\ds{m_1,\alpha_1} 
\cdots\ds{m_{i-1},\alpha_{i-1}}
\ds{m_{i}-1,1+\alpha_i}\ds{m_{i+1},\alpha_{i+1}}
 \cdots\ds{m_n,\alpha_n}}{g,d} \nonumber\\&& \ \ \ +\,  d
	\ccor{\ds{m_1,\alpha_1} \cdots 
\ds{m_n,\alpha_n}}{g,d}.
	\earray
\item {\bf Dilaton Axiom.} For either $n+2g \geq 3$ or $d>1,n\geq
1$, 
	\beq
	\ccor{\ds{1,0}\ds{m_1,\alpha_1} \cdots
\ds{m_n,\alpha_n}}{g,d} = (2g-2 + n) \ccor{\ds{m_1,\alpha_1} \cdots
\ds{m_n,\alpha_n}}{g,d}.
	\eeq
\end{itemize} 
In degree zero, there are the following exceptional cases:
	\barray
	\ccor{\ds{0,0}\ds{0,a}\ds{0,b}}{0,0} &=& \int_{\P^1} \gamma_a
	\cup \gamma_b \nonumber\\
	\ccor{\ds{0,1}}{1,0} &=&  -\oover{24} \nonumber\\
	\ccor{\ds{1,0}}{1,0} &=& \oover{12}\nonumber
	\earray
The dimension of the virtual fundamental class is
	\[
	\mbox{vdim} (\overline{\cM}_{g,n} (\P^1, d)) = 2 (g-1) + 2d + n \ .
	\]
Thus, from the degree axiom, we see that the non-vanishing GW-invariants are of the form 
	\[
	\ccor{ \ds{n_1,0} \cdots \ds{n_k,0} \, \ds{m_1,1}
	\cdots \ds{m_{\ell},1}}{g,d} 
	\]
where
	\[
	2 (g-1) + 2d + k = \sum^k_{i=1} n_i + \sum^{\ell}_{i=1} m_i \, .
	\]
In particular, the only non-vanishing GW-invariants
that do not contain the tautological $\psi$-classes are:
	\[
	\ccor{\ds{0,1}\ds{0,1} \cdots \ds{0,1}}{0,1} =1 \ \ \mbox{ and } \ \
	\ccor{\ds{0,0}\ds{0,0}\ds{0,1}}{0,0} =1
	\]
in genus-0 and
	\[
	\ccor{\ds{0,1}}{1,0} = -\frac{1}{24}
	\]
in genus-1.  All other non-vanishing GW-invariants thus contain the
	tautological $\psi$-classes, and we call them descendant
	GW-invariants.

%%%%%%%%%%%%%%%%%%%
%  computations
%%%%%%%%%%%%%%%%%%%
\subsection{Genus-Zero} \label{subsec:g=0 cor}
Topological recursion relations (TRRs) generally follow from the equality of
the tautological classes with boundary classes on the moduli space
$\cM_{g,n}$, and  they are used to reduce the degree of the descendants
inside 
correlators. 
The TRRs for the generating
functions of  the GW-invariants
of $\P^1$ are given  by
	\barray  \label{eq:TRR g=0}
	\cor{\ds{m_1,\alpha_1} \ds{m_2,\alpha_2} \ds{m_3,\alpha_3}}{0}
	&=& \cor{\ds{m_1-1,\alpha_1} \ds{0,0}}{0}\cor{\ds{0,1}
	\ds{m_2,\alpha_2} \ds{m_3,\alpha_3}}{0}\nonumber\\
	&&\hspace{5mm} +\ \cor{\ds{m_1-1,\alpha_1} \ds{0,1}}{0}\cor{\ds{0,0}
	\ds{m_2,\alpha_2} \ds{m_3,\alpha_3}}{0} \ .
	\earray
We can compute the $n$-point descendant GW-invariants by using this relation and
other topological ``axioms;'' the  numbers that we compute are thus
independent of the Virasoro conjecture.  
We will later use these
information to test numerically the Virasoro conjecture in genus-2.

We start with  the two-point functions $\ccor{\ds{m_1,\alpha_1}
\ds{m_2,\alpha_2}}{0}$ by noticing that there are two ways of
reducing the invariants
$\ccor{\ds{m_1,\alpha_1} \ds{m_2,\alpha_2} \ds{0,0}}{0}$.  That is, we
can either use the genus-0 TRR \eqr{eq:TRR g=0} to reduce the degree
of $\ds{m_1,\alpha_1}$ or use the string
equation to reduce the invariant into a sum of two-point functions:
	{\small \beq \label{eq:g=0 2-pt recusions}
	\ccor{\ds{m_1-1,\alpha_1} \ds{m_2,\alpha_2} }{0}
+\ccor{\ds{m_1,\alpha_1} \ds{m_2-1,\alpha_2} }{0} =
\ccor{\ds{m_1-1,\alpha_1} \ds{0,0} }{0}  \ccor{\ds{0,1}
\ds{m_2,\alpha_2} }{0} + \ccor{\ds{m_1-1,\alpha_1} \ds{0,1} }{0}
\ccor{\ds{0,0}
\ds{m_2,\alpha_2} }{0}
	\eeq}

\vspace{-2mm}
\noindent
The two-point functions of the form $ \ccor{\ds{m,\alpha} \ds{0,\beta}
}{0}$ can be computed by using the TRR of Eguchi-Hori-Xiong
\eqr{eq:TRR-EHX} and are 
given by \eqr{eq:app g=0 2pt}.
Now, \eqr{eq:g=0 2-pt recusions} gives us a
set of recursive relations among the two-point functions.  For example,
define
\beq
 X(m) = \ccor{\ds{2m-1,0} \ds{2d-2m+1,0}}{0,d}
 \ \ \ \mbox{ and }\  \ \  Y(m) = \ccor{\ds{2m,0}
\ds{2d-2m,0}}{0,d} \ .
\eeq	
Then, we obtain the relations
	\beq
	Y(m) = -X(m) + A(m,d) \hspace{.6cm}\mbox{and}\hspace{.6cm}
	X(m) = -Y(m-1) + A(d-m+1,d)\, ,
	\eeq
where	
	\beq
A(m,d) = \frac{(-2 c_{m-1} - 1/m) (-2 c_{d-m})}{m
(m-1)!^2 (d-m)!^2} 
	\eeq
and
	\beq
 c_m =\sum^m_{k=1} \,\oover{k}\ .
	\eeq
We can solve for $X(m)$ and $Y(m)$ recursively and obtain
	\barray
	 \ccor{\ds{2m-1,0} \ds{2d-2m+1,0}}{0,d}\, =\, X(m) &=&
 2\, \frac{c_d}{{{d!}^2}} + \sum^m_{k=2} \widetilde{\Delta_1}(k,d)\,
,\nonumber\\ 
\ccor{\ds{2m,0} \ds{2d-2m,0}}{0,d} \,= \,Y(m)  &=&  -2\, \frac{c_d}{{{d!}^2}} + \sum^m_{k=1}
\Delta_1(k,d) \, , 
	\earray
where
	\barray
	\Delta_1(k,d) &=& A(k,d) - A(d-k+1,d)\, , \nonumber\\ 
	\widetilde{\Delta_1}(k,d) &=& A(d-k+1,d) - A(k-1,d)\, ,
	\earray

\noindent
and the summation is set to zero whenever the lower limit exceeds the upper limit.
Other two-point functions are similarly determined, and we summarize
the results in Appendix \ref{sec:appendix-genus0}.   The one-point
descendants can be obtained from the two-point functions by using the
string equation.

The genus-0 TRR for the three-point GW-invariants are
	\beq
	\ccor{\ds{m_1,\al_1} \ds{m_2,\al_2} 
	\ds{m_3,\al_3}}{0,d} =\left\{ \begin{array}{ll}
	\ccor{\ds{m_{1}-1,\al_1}\ds{0,0} }{0,d'} \ccor{\ds{0,1} \,\ds{m_2,\al_2} 
	\,\ds{m_3,\al_3}}{0,d''} &\mbox{if $m_1 +\al_1$ is odd,} \\
	\ccor{\ds{m_{1}-1,\al_1} \ds{0,1}}{0,d'+\onehalf} \ccor{\ds{0,0} \,\ds{m_2,\al_2} 
	\,\ds{m_3,\al_3}}{0,d''-\onehalf} & \mbox{if $m_1 + \al_1$ is even.}
	\end{array}\right.\eeq
where the degree $d$ of the holomorphic maps must satisfy
	\barray
	2d+1 &=& \sum^3_{i=1} m_i + \sum^3_{i=1} \al_i\, , \nonumber\\
	d' &=&\frac{m_1+\al_1 -1}{2} \ \ \mbox{ and }\ \ d''=d-d' \ .
	\earray
By the divisor axiom and the string equation, we can further
manipulate the above quantities to produce two-point
functions.  For example, we have
	\beq
	\ccor{\ds{0,1} \,\ds{m_2,\al_2} \ds{m_3,\al_3}}{0} = d'' \ccor{\ds{m_2,\al_2} 
	\,\ds{m_3,\al_3}}{0} +  \ccor{\ds{m_2-1,\al_2+1} 
	\,\ds{m_3,\al_3}}{0} +  \ccor{\ds{m_2,\al_2} 
	\,\ds{m_3-1,\al_3+1}}{0} \, .
	\eeq
Hence, using the previous computations of the two-point
invariants, we can now also compute arbitrary  three-point functions,
	whose closed-form expressions are possible but not
	instructive.  We thus compute them numerically using
a computer program and
tabulate some of the invariants in Appendix \ref{sec:lists g=0}.

Similarly and unfortunately, we have reduced the problem of computing
the higher-point invariants 
to an exercise in computer programming.  We differentiate the TRR
\eqr{eq:TRR g=0} repeatedly and use the divisor, dilaton, and
string equations to reduce the number and degrees of the fields
appearing in the right-hand side of equation.  Using this simple but
tedious approach, we have completed a computer program which
computes up to 7-point functions of arbitrary degrees and have
included a
few
examples in Appendix \ref{sec:app-lists}.

%%%%%%%%%%%%%%%%%
%  Genus-One
%%%%%%%%%%%%%%%%%

\subsection{Genus-One} \label{subsec:g-1 GW}
Because there exists another TRR \cite{Witten} in genus-1, it is also
possible to compute the genus-1 descendant GW-invariants.  For $\P^1$,
the relation has the following simple form:
	\beq	\label{eq:genus-1TRR}
	\cor{\ds{n,\alpha}}{1} = \cor{\ds{n-1,\alpha} \ds{0,0}}{0}
	\cor{\ds{0,1}}{1} + \cor{\ds{n-1,\alpha} \ds{0,1}}{0}
	\cor{\ds{0,0}}{1} + \oover{12} \cor{\ds{n-1,\alpha} \ds{0,0}
	\ds{0,1}}{0} \ .
	\eeq
Setting all $t^{\alpha}_m=0$, \eqr{eq:genus-1TRR} yields
	\barray
	\ccor{\ds{2d+1,0}}{1,d} &=&\left\{\begin{array}{ll} 
	\oover{12\, d!^2}\, (c_d - 2d 
c_{d-1} -1 )\ &  \mbox{for } d\neq 0\, ,\\ \oover{12}\ &  \mbox{for }
d=0\, ,\  \mbox{ and} \end{array} \right.\nonumber\\
	\ccor{\ds{2d,1}}{1,d} &=& -\frac{1-2d}{24 \, d!^2} \, .
	\earray

To compute the two-point functions\footnote{There are of course a few
obvious ones that one can compute by using the divisor and string
equations.  For example, one can show that $\ccor{\ds{0,1}\ds{0,1}}{1}
=0,\, \ccor{\ds{1,0}\ds{0,1}}{1}= -1/24,\, \ccor{\ds{1,0}\ds{1,0}}{1}= 1/12.$}, we differentiate
\eqr{eq:genus-1TRR} with respect to a descendant variable
$t^{\beta}_m$
 and get an
equation which is valid in the large phase space:
	{\small \barray
\cor{\ds{n,\alpha}\ds{m,\beta}}{1} &=& \cor{\ds{n-1,\alpha}\ds{m,\beta} \ds{0,0}}{0}
	\cor{\ds{0,1}}{1} + \cor{\ds{n-1,\alpha}\ds{0,0}}{0}
	\cor{\ds{0,1}\ds{m,\beta}}{1} +
\cor{\ds{n-1,\alpha}\ds{m,\beta} \ds{0,1}}{0}
	\cor{\ds{0,0}}{1} \nonumber \\ && + \ \cor{\ds{n-1,\alpha}\ds{0,1}}{0}
	\cor{\ds{0,0}\ds{m,\beta}}{1}  + \oover{12} \cor{\ds{n-1,\alpha}\ds{m,\beta} \ds{0,0}
	\ds{0,1}}{0}\ .
	\earray}
At zero couplings, it becomes
	{\small \barray
\ccor{\ds{n,\alpha}\ds{m,\beta}}{1} &=& -\oover{24} \ccor{\ds{n-1,\alpha}\ds{m,\beta} \ds{0,0}}{0}
	 + \ccor{\ds{n-2,\alpha}}{0}
	\ccor{\ds{0,1}\ds{m,\beta}}{1} \nonumber \\ && +\  \ccor{\ds{n-1,\alpha}\ds{0,1}}{0}
	\ccor{\ds{m-1,\beta}}{1}  + \oover{12} \ccor{\ds{n-1,\alpha}\ds{m,\beta} \ds{0,0}
	\ds{0,1}}{0}\ .
	\earray}

\vspace{-2mm}
\noindent
which contains only genus-0 invariants and genus-1 one-point
functions, upon using the divisor axiom.  Closed-form answers 
for these two-point functions are again possible, but they are not perhaps so
illuminating.  We thus omit the explicit expressions in the paper but list some of
their values in  Appendix \ref{sec:app-genus1}.   

It is clear that the higher-point genus-1 descendant GW-invariants can be
similarly computed in terms of the genus-0 and the lower-point genus-1 invariants 
by repeatedly differentiating \eqr{eq:genus-1TRR}.   We have again
implemented the algorithm into a computer program which computes up to
5-point functions of arbitrary degrees, and we have also tabulated
a collection of our results up to 3-point functions in Appendix
\ref{sec:app-genus1}.

%%%%%%%%%%%%%%%%%
%  Genus 2
%%%%%%%%%%%%%%%%%

\subsection{Genus-Two}
Combined with our previous computations, 
we now use a genus-2 TRR \cite{Getzler2} to compute
up to 3-point descendant GW-invariants in arbitrary degrees. 
 Getzler's TRR \eqr{eq:app
genus-2} directly leads to the one-point functions, and their
derivatives yield the desired two- and three-point
functions  (See
Appendix \ref{app:list g=2}).  His TRRs for two-point functions, on the
other hand, seem somewhat mysterious to us, and we were not able to
produce the GW-invariants which satisfy the topological axioms and
the genus-2 Virasoro constraints.

%%%%%%%%%%%%%
%  Hodge integrals
%%%%%%%%%%%%%

\subsection{Hodge Integrals}
This section marks the end
of our torture with programming.  

Let $\pi,\Sigma$ and $\omega$ be as in \S\ref{sec:properties}.  The Hodge
bundle $E=\pi_* \omega$ over $ \overline{\cM}_{g,n}(V,\beta)$ 
is a rank-$g$ sheaf of holomorphic sections of $H^0(\Sigma,
\omega_\Sigma)$, where $ \omega_\Sigma$ is 
the canonical sheaf of $\Sigma$.  The $\lambda_i$ classes are defined
to be the $i$-th Chern classes of the Hodge bundle, and a generalization
of the Gromov-Witten integral of the form \eqr{eq:GW-integral}
including the $\lambda$-classes is called a Hodge integral.
In \cite{FP}, Faber and Pandharipande have found a set of differential
operators that annihilate the generating function for 
 Hodge integrals.  In principle, their theorem allows one to compute
the Hodge integrals on the moduli space of stable maps
in terms of the descendant Gromov-Witten invariants.  In practice,
however, it is difficult to compute the GW-potential in the large phase
space, and it is precisely for this reason that some kind of an
integrable structure such as the Virasoro constraints is desirable in
studying the intersection theory.  It is, however, often the case that
the Virasoro constraints alone are not strong enough to determine
the GW-invariants on non-trivial target spaces.  In this paper, 
we have taken a
different approach to computing the invariants, and 
we have seen that for
$\P^1$, the known topological recursion relations  allow
one to compute all the GW-invariants up to genus-2.  
Using these results, the work of Faber and Pandharipande
completely determines all the Hodge integrals for $\P^1$ up to
genus-2. 

The expressions of the differential equations for
$\P^1$ are particularly simple:
{\small	\barray 
	\cor{ch_{2\ell-1}(E)}{g} &=& \frac{B_{2\ell}}{(2\ell )!} \left[
	\cor{\ds{2\ell,0}}{g} - \sum^{\infty}_{m} t^0_m\,\cor{\ds{m+2\ell
-1,0}}{g}  - \sum^{\infty}_{m} t^1_m \,\cor{\ds{m+2\ell -1,1}}{g}
\right. \nonumber\\
	&& \left.+ \sum^{2\ell -2}_{m=0} (-1)^m \cor{\ds{m,0} \ds{2\ell
-2-m,1}}{g-1} + \sum_{g'+g''=g}\, \sum^{2\ell -2}_{m=0} (-1)^m
\cor{\ds{m,0}}{g'}\cor{ \ds{2\ell-2-m,1}}{g''}\right]\, ,\nonumber
	\earray}

\noindent
where $B_{2\ell}$ are the Bernoulli numbers.  

The first non-trivial Hodge integral is $\ccor{\ds{0,0}\,
\lambda_1}{1,0}$, which can be computed explicitly as follows:
	\barray\label{eq:faber-pand}
	\ccor{\ds{0,0}\, \lambda_1}{1,0} &=& \int_{\P^1 \times
\overline{\cM}_{1,1}} \lambda_1 ( c_1(\P^1) - \lambda_1 ) \nonumber\\
	&=&   \int_{\P^1 \times
\overline{\cM}_{1,1}} c_1(\P^1)\, \lambda_1\nonumber\\
	&=& 2 \int_{\overline{\cM}_{1,1}} \,\lambda_1\nonumber\\
	&=& \oover{12}\, ,
		\earray
where we have used the
formula for the Euler class of the obstruction bundle from \cite{KM2},
Mumford's relation $\lambda_g \lambda_g =0$,
and the numerical value of $\int_{\overline{\cM}_{1,1}} \,\lambda_1 =
1/24$ from \cite{FP}.   As simple illustrations, we have implemented
our computer program to compute
this and other arbitrary Hodge
integrals involving up to two-descendants in genus-1. ( See Appendix
\ref{sec:app-genus1} for a partial list.)  The
genus-2 cases are similarly treated:  The $\ell=1$ relations in
\eqr{eq:faber-pand} lead to $\lambda_1$ Hodge integrals, and the
Mumford's relation $2 \lambda_2 =  \lambda^2_{1}$ yields the
$\lambda_2$ integrals.  Since the algorithm is obvious by now, we do
not explicitly carry out the computations.

%%%%%%%%%%%%%########################################
%    Virasoro Constraints
%%%%%%%%%%%%%########################################

\setequation
\section{Virasoro Constraints}\label{sec:Virasoro}
Let $Z$ be the generating function\footnote{$Z$ is called the partition
function in the physics literature.} for GW-invariants:
	\beq
	Z = \exp \left[ \sum_g \lambda^{2g-2}\ccor{\exp\left(\sum_{m,a} t^a_m \ds{m,a} 
\right)}{g}\right] \, ,
	\eeq
and define $z_{n,g}$ to be the genus-$g$ contribution\footnote{That
is, the  coefficient of $\lambda^{2g-2}$ in  $Z^{-1} L_n Z$.}  to $Z^{-1} L_n Z$.
The Virasoro constraints for $\P^1$ are
{\small
\barray
z_{n,g} &= & 0 \nonumber\\
	&=&\sum^{\infty}_{m=0}\left[ \alpha (m,n)\, t^0_m\,
	\cor{\ds{n+m,0}}{g} + 2\,
	\beta(m,n)\, t^0_m\, \cor{\ds{n+m-1,1} }{g} +
	\gamma(m,n)\, t^1_m\, \cor{\ds{n+m,1}}{g}\right]\nonumber\\ 
	&&-\alpha(1,n)\, \cor{\ds{n+1,0} }{g} 
	-2\,
	\beta(1,n) \cor{\ds{n,1} }{g}\nonumber\\
	&&+ \sum^{n-2}_{m=0}  \delta(m,n)\,
	\left[ \cor{\ds{m,1} \, \ds{n-m-2,1}}{g-1} 
	+ \sum_{g'+g'' =g}  
	\cor{\ds{m,1}}{g'}\cor{\ds{n-m-2,1}}{g''} 
	 \right]\, , \nonumber\\
\earray}
where we have assumed that $n>0$ and the constants are given by
	\barray
	\alpha(m,n) &=&  m\, \left( \frac{(n+m)!}{m!}\right)\, , \nonumber\\
	\beta(m,n) &=& \frac{(n+m)!}{m!} \left[1  + m (c_{n+m} -
	c_m)\right]\, , \nonumber\\
	\gamma(m,n) &=& \frac{(n+m+1)!}{m!}\, ,\nonumber\\
	\delta(m,n) &=& (m+1)!\, (n-m-1)!\ .
	\earray

\subsection{Genus-Zero}
The genus-0 Virasoro constraints are
{\small
\barray
0\, =\, z_{n,0} &=& \sum^{\infty}_{m=0}\left[ \alpha (m,n)\, t^0_m\,
	\cor{\ds{n+m,0} }{0} + 2\,
	\beta(m,n)\, t^0_m\, \cor{\ds{n+m-1,1} }{0} +
	\gamma(m,n)\, t^1_m\, \cor{\ds{n+m,1}
	}{0}\right]\nonumber\\ 
	&&-\alpha(1,n)\, \cor{\ds{n+1,0} }{0} 
	-2\,
	\beta(1,n) \cor{\ds{n,1} }{0}+
	\sum^{n-2}_{m=0}
	\delta(m,n)\,
	\cor{\ds{m,1}	}{0}\,\cor{\ds{n-m-2,1}}{0}  \ . \label{eq:z_{n,0}} 
\earray}

\vspace{-3mm}
\noindent
Taking derivatives of \eqr{eq:z_{n,0}} with respect to the
variables $t^{\alpha}_m$ yields a set of equations which the
GW-invariants must satisfy.  Let $\cI = \{ 1,2,\ldots k\}$ and $\cJ =
\{ 1,2,\ldots \ell\}$ be two index sets, and $\cI', \cI'', \cJ'$, and
$\cJ''$ their partitions into two complementary subsets. The set $\cI$
labels the descendants of the identity and $\cJ$ those of the
hyperplane class. Then,  one finds

	{\small
	\barray 	\label{eq:g=0 constraint}
	0&=&\sum^k_{i=1} \alpha(m_i,n)\, \ccor{ \ds{n+m_i,0} (\prod_{j\neq i}
\ds{m_j,0}) \ds{s_1,1} \cdots \ds{s_{\ell},1}}{0}  
	- \alpha(1,n) \ccor{ \ds{n+1,0} \ds{m_1,0} \cdots
	\ds{m_{k},0}  \ds{s_1,1} \cdots \ds{s_{\ell},1}}{0} \nonumber\\ 
	&&+\, 2\sum^k_{i=1} \beta(m_i,n)\, \ccor{ \ds{n-1+m_i,1} (\prod_{j\neq i}
\ds{m_j,0}) \ds{s_1,1} \cdots \ds{s_{\ell},1}}{0} -2 \beta(1,n) \ccor{ \ds{n,1} \ds{m_1,0} \cdots
	\ds{m_{k},0}  \ds{s_1,1} \cdots \ds{s_{\ell},1}}{0}  \nonumber\\
	&&+\sum^{\ell}_{a=1} \gamma(s_a,n)\, \ccor{  \ds{m_1,0} \cdots
	\ds{m_{k},0}  \ds{n+s_a,1}  (\prod_{b\neq a} \ds{s_b,1})}{0}
\nonumber\\
	&& \sum^{n-2}_{q=0} \sum_{\cI',\cI'',\cJ',\cJ''} \delta(q,n) 
	\ccor{\ds{q,1} (\prod_{i\in\cI'} \ds{m_i,0}) \, (\prod_{a\in\cJ'}
\ds{s_a,1})}{0} \ccor{\ds{n-q-2,1} (\prod_{j\in\cI''} \ds{m_j,0}) \, (\prod_{b\in\cJ''}
\ds{s_b,1})}{0} \, .
	\earray}

\vspace{-2mm}
\noindent
The Virasoro constraints are actually proven to hold in genus-0
\cite{Tian}, and we have numerically checked that the constraints
\eqr{eq:g=0 constraint} are indeed satisfied for roughly 5000 cases
containing up to four-point functions.  This test makes it fairly
certain that our
computer generated answers of the genus-0 GW-invariants are correct.

%%%%%%%%%%%%%%%%%%%
%
%%%%%%%%%%%%%%%%%%%

\subsection{Genus-One}
Since $\P^1$ has a semi-simple quantum cohomology,  the Virasoro
conjecture is true also in genus-1 \cite{DZ,Liu}.
In this case, the Virasoro constraints take the form 
{\small
\barray
0\, =\, z_{n,1} &=& \sum^{\infty}_{m=0}\left[ \alpha(m,n)\, t^0_m\,
	\cor{\ds{n+m,0}}{1} + 2\,
	\beta(m,n)\, t^0_m\, \cor{\ds{n+m-1,1} }{1} +
	\gamma(m,n)\, t^1_m\, \cor{\ds{n+m,1}}{1}\right]\nonumber\\ 
	&&- \ \alpha(1,n)\, \cor{\ds{n+1,0} }{1} 
	-2\,
	\beta(1,n)\, \cor{\ds{n,1} }{1}\nonumber\\
	&&+ \sum^{n-2}_{m=0}  \delta(m,n)\,
	\left[ \cor{\ds{m,1} \, \ds{n-m-2,1}}{0} 
	+ 
	2\cor{\ds{m,1}}{0}\cor{\ds{n-m-2,1}}{1}   
	 \right] \ ,  
\earray}

\vspace{-2mm}
\noindent
and they yield constraints that are similar to \eqr{eq:g=0 constraint}.
Using the genus-0 and genus-1 descendant invariants that were computed 
 in \S\ref{subsec:g=0 cor} and \S\ref{subsec:g-1 GW}, we have checked that
 over 7000
Virasoro constraints which involve up to 4-point genus-1 GW-invariants
are satisfied.

\subsection{Genus-Two}
The status of the Virasoro constraints in genus-2 is still
conjectural, and it would be interesting to see if the
GW-invariants which are obtained from either rigorously
derived  TRRs or  algebraic geometry actually satisfy the Virasoro
constraints in this case and in higher genera.  

We have checked that our results
are indeed 
consistent with about 1100 Virasoro constraints containing up to
3-point genus-2 invariants.  As previously mentioned, the genus-2
descendant invariants obtained from Getzler's TRRs for 2-point
functions (equation (7) in \cite{Getzler2}, or his corrected version
\cite{Getzler3}) do not seem to satisfy the Virasoro constraints, whereas
the invariants obtained from his 1-point TRRs (equation (6)
in \cite{Getzler2})
do satisfy the
Virasoro constrains as well as the required topological axioms. 
We do not understand the origin of our, or possibly his, mistake.

%%%%%%%%%%%%%%%%%%#######################################################
%  Higher Genus and Localization
%%%%%%%%%%%%%%%%%%#######################################################

\subsection{Speculations on Higher-Genus Cases: TRRs and Localizations}
We find that the Virasoro constraints by themselves do not provide an
efficient computational tool unless we already know many of the
GW-invariants that are to be used in the constraint equations.  In the
pure gravity case, the Virasoro constraints relate  $\ds{n+1,0}$ with
 $\ds{n,0}$, thus providing an
effective recursions among the descendants.
In the $\P^1$ case, however, the  Virasoro constraints relate
$\ds{n+1,0}$  with $\ds{n,1}$, but there is no relation between
$\ds{n,1}$ and $\ds{n-1,0}$.  This pattern of recursion explains why
the Virasoro constraints generally cannot determine the GW-invariants
by themselves.  

Motivated by the previous computations, 
it is tempting to speculate that there may exist higher-genus TRRs that
completely determine the GW-invariants.  The only higher-genus TRRs
that are known 
to us so far are those found by Eguchi and Xiong in \cite{EX}.  
Unfortunately, their derivation crucially depends on the assumption that the
genus-$g$ free energy $\cF_g (t)= \ccor{\exp\left(\sum_{m,a} t^a_m \ds{m,a} 
\right)}{g}$ is a function of genus-0 correlation functions
in the large phase space.
That is,  their derivation assumes that 
	\beq \label{eq:assumption}
	\cF_g (t) = \cF_g ( u_{\alpha_{1}}(t), u_{\alpha_1\alpha_2}(t),
\ldots, u_{\alpha_1\alpha_2\cdots\alpha_{3g-1}}(t))
	\eeq
where $u_{\alpha_1\alpha_2\cdots\alpha_{k}} :=
\cor{\ds{0,0}\ds{0,\alpha_1}\cdots \ds{0,\alpha_k}}{0} =
\partial^{k+1} \cF_0/\partial t^0_0 \cdots\partial t^{\alpha_k}_0$.
At first sight, it tells us that a genus-$g$ GW-invariant can be
expressed in terms of genus-$0$ invariants; more precisely, it
determines the functional dependence of the free energy on the
variables $t^{\alpha}_m$ through the genus-zero quantities
$   u_{\alpha_{1}}(t), u_{\alpha_1\alpha_2}(t),
\ldots, u_{\alpha_1\alpha_2\cdots\alpha_{3g-1}}(t).$ 
In the rest of this section, we
use the technique of 
localization\footnote{We are grateful to Prof. Tian for suggesting
this analysis.
We are ineluctably led to make it absolutely
clear at this point that we do not have a satisfactory understanding
of the ideas involving localizations and that the ensuing statements are
only speculative.  
As we do not feel competent enough to present a rigorous proof, we are
somewhat reluctant to present our arguments here.  Nevertheless, with the
hope that our honesty would engender further objectivity and caution
from the readers than they would normally require, we proceed.} to
comment on the validity of this assumption for complex projective
spaces $\P^r$
admitting torus actions.  

We will be very brief and use the results of \cite{GrP,K2}.
Given a compact complex projective variety 
$V$ and a holomorphic vector bundle
$E\ra V$, equipped with a torus action $T\simeq\C^*\times\cdots\times
\C^*$ on $(V,E)$, the Atiyah-Bott fixed points formula reduces the
integrals 
of characteristic classes of $E$ over $V$ to new integrals over fixed
loci of the torus action on $V$ \cite{AB,Bott}.  Recall that the
GW-invariants are defined 
to be integrals of certain characteristic classes over the virtual
fundamental class $[\overline{\cM}_{g,n} (V,\beta)]^{\mbox{vir}}$.
  The torus action on $V$
can be naturally lifted to $\overline{\cM}_{g,n} (V,\beta)$ by translating the 
stable maps.     The work of Graber and Pandharipande \cite{GrP} states
that for non-singular projective varieties $V$, there exists a
localization formula for the virtual fundamental class $[\overline{\cM}_{g,n}
(V,\beta)]^{\mbox{vir}}$, and thus the associated GW-invariants can be
defined by integrals over the virtual classes of the fixed loci of the
torus action on the moduli space.  In particular, the localization
formula holds for projective spaces $\P^r$, and the final result
which we need is that an arbitrary
GW-invariant of $\P^r$ can be expressed as a
sum of Hodge integrals over products of the moduli
spaces $\overline{\cM}_{g',n'}$ of pointed Riemann surfaces; that is, roughly
	\beq\label{eq:localization}
 	\ccor{\ds{m_1,\alpha_1} \cdots\ds{m_n,\alpha_n}}{g,d} = \sum_{\Gamma}
\int_{ \overline{\cM}_{\Gamma}} \psi^{m_1}_1 \cdots \psi^{m_n}_n\,
\frac{\mbox{weights}}{e(N^{\mbox{\tiny vir}}_{\Gamma})} 
	\eeq
where the sum is over all the  fixed loci represented by certain
``graphs'' $\Gamma$, $e(N^{\mbox{\tiny vir}}_{\Gamma})$ is the Euler
class of the virtual normal bundle to $\overline{\cM}_{\Gamma}$,
$\psi_i$ the pull-back of the first Chern class of the cotangent
bundle at $i$-th marked point on the Riemann surface,  and
the ``weights'' are determined by the torus action and on the
cohomology classes $\gamma_{\alpha_j}\in H^*(\P^r,\C)$.  Furthermore,
the fixed loci represented by the graph $\Gamma$
 are products of the moduli spaces of pointed stable curves:
	\beq
	\overline{\cM}_{\Gamma} = \prod_{\mbox{\tiny vertices}}
\overline{\cM}_{g(v),\mbox{\tiny val}(v)} \ ,
	\eeq  
with $g(v)\leq g$ representing the arithematic genus of the contracted
component of the domain curve.
We refer the
reader to \cite{GrP,K2} for the specific definitions of the notations
which are actually not so essential for our discussion.  

The Euler class $e(N^{\mbox{\tiny vir}}_{\Gamma})$  of the virtual
normal bundle introduces the $\lambda$-classes, and the resulting
Hodge integrals can be reduced to pure $\psi$ integrals by using
Faber's algorithm \cite{Faber}.
Now, we recall the fact that for the intersection theory of the
tautological divisors on the moduli
space of stable pointed curves, the genus-$g$ free energy, for $g>0$,
 is actually a
function of the genus-0 correlators \cite{EYY}:
	\beq \label{eq:EYY}
	\cF_g(t)_{\mbox{\tiny point}} = \cF_g ( u^{(1)}(t), \cdots,
u^{(3g-2)}(t))_{\mbox{\tiny point}} \, ,
	\eeq
where $u := \cor{\ds{0}\ds{0}}{0} = \partial^2 \cF_0/\partial
t_0\, \partial t_0 $ and $u^{(i)} =\partial^i u/\partial
{t_0}^i $.  Combined with the localization formula
\eqr{eq:localization},  
the form of \eqr{eq:EYY} implies that indeed each GW-invariant can be
expressed in terms of $  u_{\alpha_{1}}(0), u_{\alpha_1\alpha_2}(0),
\ldots, u_{\alpha_1\alpha_2\cdots\alpha_{3g-1}}(0)$ and the values of
their derivatives at the origin of the phase space.  This statement
is however much weaker than the assumption \eqr{eq:assumption}.  That
is, our analysis does not show that the functional dependence of
$\cF_g(t)$ on $t^{\alpha}_m$ is only
through  $u_{\alpha_{1}}(t), u_{\alpha_1\alpha_2}(t),
\ldots, u_{\alpha_1\alpha_2\cdots\alpha_{3g-1}}(t)$.  Even though we
cannot prove the statement at the moment, we believe that our
approach deserves a further consideration.

%%%%%%%%%%%%%%%%%%########################################################
%  Hurwitz Numbers
%%%%%%%%%%%%%%%%%%########################################################
\setequation
\section{Recursion Relations for Simple Hurwitz Numbers} \label{sec:Hurwitz}
Hurwitz numbers, whose study had been initiated by Hurwitz more than a
century ago \cite{Hurwitz}, count the number of inequivalent ramified
coverings of 
a sphere by Riemann surfaces with specified branching conditions over
one point called $\infty$.  The original approach of Hurwitz relates
the problem to transitive factorizations of permutations into
transpositions.  Recently, new insights have been gained from 
developments in the absolute and relative Gromov-Witten theory 
\cite{GJV,LZZ,Vakil}.  

In this section, we take a very modest goal of obtaining new recursion
relations for the genus-0
and genus-1  
simple Hurwitz numbers which enumerate the coverings with
no ramification over $\infty$.  We take two different approaches which
yield similar but inequivalent recursion relations.  In the notations
of the previous sections, the  genus-$g$ simple Hurwitz numbers are
defined\footnote{We are grateful to R. Vakil for 
this definition.} by the descendant GW-invariants of $\P^1$ as
	\beq
	H^g_d := \ccor{\ds{1,1}^{2d+2g-2}}{g}\ .
	\eeq 
We first show that the genus-0 and genus-1 TRRs immediately lead to
relations among the simple Hurwitz numbers in these genera.  
We then use the Virasoro constraints to derive new relations which
could be generalized to higher genera.

\subsection{From Topological Recursion Relations}
In this subsection, by using the TRRs \eqr{eq:TRR g=0} and \eqr{eq:genus-1TRR},
we derive a new recursion relation for simple
Hurwitz numbers in genus-0 and reproduce the known result of Graber
and Pandharipande in genus-1.

\begin{claim}  The genus-0 simple Hurwitz numbers satisfy
\beq \label{eq:g=0 hurwitz}
	H^0_d = 4 \sum^{d-3}_{k=0} \left(\hspace{-1.5mm}\begin{array}{c} 
			2d-5\\ 2k \end{array}\hspace{-1.5mm} \right) (d-k-1)(d-k-2)
	(k+1)^2 H^0_{k+1} \, H^0_{d-k-1}.
\eeq
\end{claim}

\noindent
{\sc Proof:}  We need the following equations which are implied
by the string and the divisor equations:
	\barray \label{eq:hur-rec}
	\ccor{\ds{0,1}\ds{0,0}\ds{1,1}^{2k-1}}{0} &=& (2k-1)\,k^2\,
\ccor{\ds{1,1}^{2k-2}}{0} \nonumber\\
	\ccor{\ds{0,1}\ds{0,1}\ds{1,1}^{2k}}{0} &=& (k+1)^2\,
\ccor{\ds{1,1}^{2k}}{0} ,	
	\earray
and similarly for $\ccor{\ds{0,0}\ds{1,1}^{2k-1}}{0}$ and
$\ccor{\ds{0,1}\ds{1,1}^{2k}}{0}$. 
Differentiating the genus-0 TRR \eqr{eq:TRR g=0} yields
	\barray
	\ccor{\ds{1,1}^{2n}}{0} &=& \sum^{2n-3}_{\ell=0}
\left(\hspace{-1.5mm}\begin{array}{c}  
			2n-3\\ \ell \end{array}\hspace{-1.5mm}\right) \left[
\ccor{\ds{0,1}\ds{0,0} \ds{1,1}^{\ell}}{0}
\ccor{\ds{0,1}\ds{1,1}^{2n-\ell-1}}{0} + \ccor{\ds{0,1}\ds{0,1} \ds{1,1}^{\ell}}{0}
\ccor{\ds{0,0}\ds{1,1}^{2n-\ell-1}}{0}\right]\nonumber\\
	&=& \sum^{n-1}_{k=1} \left(\hspace{-1.5mm}\begin{array}{c} 
			2n-3\\ 2k-1 \end{array}\hspace{-1.5mm}\right) 
\ccor{\ds{0,1}\ds{0,0} \ds{1,1}^{2k-1}}{0}
\ccor{\ds{0,1}\ds{1,1}^{2n-2k}}{0} \nonumber\\
	&& \hspace{5mm}+\,\sum^{n-2}_{k=0} \left(\hspace{-1.5mm}\begin{array}{c} 
			2n-3\\ 2k \end{array}\hspace{-1.5mm}\right) 
\ccor{\ds{0,1}\ds{0,1} \ds{1,1}^{2k}}{0}
\ccor{\ds{0,0}\ds{1,1}^{2n-2k-1}}{0}\ ,\nonumber
	\earray
where we have used the fact that many of the correlation functions
vanish for dimensional reasons and  we have relabeled indices.  Now, 
using \eqr{eq:hur-rec} and relabeling the summation yields the
desired result. {\hfill $\qed$}

\vspace{5mm}
Together with the initial conditions $H^0_1=1$ and $H^0_2 =1/2$, these
recursion relations easily determine all the simple Hurwitz numbers in
genus-zero.  The formulae \eqr{eq:g=0 hurwitz}
are qualitatively similar to those found by
Graber and Pandharipande \cite{FanP}, but they are in fact different
recursion relations.  

Similarly, we use the genus-1 TRR to derive 
recursion relations for genus-1 Hurwitz numbers $H^1_d =
\ccor{\ds{1,1}^{2d}}{1,d}$: 
	\begin{claim} The genus-1 simple Hurwitz numbers satisfy
	\beq \label{eq:g=1 hurwitz}
		H^1_d = 2 \, \sum^{d-1}_{k=1} k^2 (d-k) (2d-2k+1) 
	\left(\hspace{-1.5mm}\begin{array}{c} 
			2d-1\\ 2k-2 \end{array}\hspace{-1.5mm} \right)
	H^0_k \, H^1_{d-k} + \oover{12} d^2(d-1)(2d-1) H^0_d 
	\eeq
	\end{claim}

\noindent
{\sc Proof:}  As in the proof of \eqr{eq:g=0 hurwitz}, we
differentiate the genus-1 TRR and use dimensional arguments to get
	\barray
	\ccor{\ds{1,1}^{2n}}{1}	&=&
	\sum^{n}_{k=1} \left(\hspace{-1.5mm}\begin{array}{c}  
			2n-1\\ 2k-1 \end{array}\hspace{-1.5mm}\right) 
	 \ccor{\ds{0,1}\ds{0,0}\ds{1,1}^{2k-1}}{0}
	\ccor{\ds{0,1} \ds{1,1}^{2n-2k}}{1}\nonumber\\
	&&\ \ \ + \sum^{n-1}_{k=0} \left(\hspace{-1.5mm}\begin{array}{c}  
			2n-1\\ 2k \end{array}\hspace{-1.5mm}\right) 
	 \ccor{\ds{0,1}\ds{0,1}\ds{1,1}^{2k}}{0}
	\ccor{\ds{0,0} \ds{1,1}^{2n-2k-1}}{1}+ \frac{2n-1}{12}
\ccor{\ds{0,1}^3\ds{1,1}^{2n-2}}{0} \nonumber
	\earray
Taking caution that $\ccor{\ds{0,1}}{1}= -1/24$, we obtain \eqr{eq:g=1
hurwitz} upon using the divisor
and the string equations. {\hfill $\qed$}

Unlike the genus-0 case, with a minor 
rearrangement of terms, it is easy to see that 
our recursion relation \eqr{eq:g=1 hurwitz} is actually equal to
that of Graber and Pandharipande.

%%%%%%%%%%%%
%
%%%%%%%%%%%%
\subsection{From Virasoro Constraints}
It is also possible to derive new recursion relations for genus-$g$ Hurwitz
numbers by combining the Virasoro constraints with some TRRs.  
Namely, the $L_1$ Virasoro constraints yield
	\beq  \label{eq:vir-hur}
	3H^g_d := 3\ccor{\ds{1,1}^{2d+2g-2}}{g}=
 3 (2d+2g-3)\, \ccor{\ds{2,1}\ds{1,1}^{2d+2g-4}}{g} -
\ccor{\ds{2,0} \ds{1,1}^{2d+2g-3}}{g}.
	\eeq
For example, for genus-0, we deduce
\begin{claim}  The genus-0 simple Hurwitz numbers $H^0_d$  satisfy the
recursion relations of the form
	{\small\barray \label{eq:vh g=0}
	H^0_d &=&(2d-3) \sum^{d-3}_{k=1}
(k+1) (d-k-1)\left[
\left(\hspace{-1.5mm}\begin{array}{c} 
			2d-6\\ 2k \end{array}\hspace{-1.5mm} \right)
	(2k+1) +
	\left(\hspace{-1.5mm}\begin{array}{c} 
			2d-6\\ 2k-1 \end{array}\hspace{-1.5mm} \right) (2d-2k-3)
\right] H^0_{k+1} H^0_{d-k-1}\nonumber\\
	&&\ - \oover{3} \sum^{d-3}_{k=1}\left(\hspace{-1.5mm}\begin{array}{c} 
			2d-5\\ 2k \end{array}\hspace{-1.5mm} \right)
	(k+1) (d-k-1) \left[ (2k-1) (2d-2k-3) + 2k(2d-2k-5)\right]
	H^0_{k+1} H^0_{d-k-1}  \nonumber\\
	&& \ + \frac{4}{3} (d-1) (2d-3) H^0_{d-1}.
\earray}
\end{claim}	

\noindent
{\sc Proof:} We need the following two recursion formulas which are
obtained from the genus-0 TRR:
{\small	\barray
	\ccor{\ds{2,1}\ds{1,1}^{2m}}{0} &=& (m+1)
\ccor{\ds{1,1}^{2m}}{0}
 +  \sum^{m-1}_{k=1}
(k+1) (m-k+1)\left[
\left(\hspace{-1.5mm}\begin{array}{c} 
			2m-2\\ 2k \end{array}\hspace{-1.5mm} \right)
	(2k+1) \right.\nonumber\\ 
	&&\left.\hspace{1cm} +\,
	\left(\hspace{-1.5mm}\begin{array}{c} 
			2m-2\\ 2k-1 \end{array}\hspace{-1.5mm} \right) (2m-2k+1)
\right] \ccor{\ds{1,1}^{2k}}{0} \ccor{\ds{1,1}^{2m-2k}}{0}\nonumber\\
	\earray}
and
{\small	\barray
	\ccor{\ds{2,0}\ds{1,1}^{2m+1}}{0} &=& -(m+1)(2m+1)
\ccor{\ds{1,1}^{2m}}{0} +  \sum^{m-1}_{k=1}
\left(\hspace{-1.5mm}\begin{array}{c} 
			2m-1\\ 2k \end{array}\hspace{-1.5mm} \right)
	(k+1) (m-k+1) \nonumber\\&&
	\left[ (2k-1)(2m-2k+1) + 2k(2m-2k-1)\right]
\ccor{\ds{1,1}^{2k}}{0} \ccor{\ds{1,1}^{2m-2k}}{0}\nonumber\\
	\earray}
The $L_1$ Virasoro constraint \eqr{eq:vir-hur} now implies our claim.
{\hfill $\qed$}

Similarly, after some algebra, one can show
\begin{claim} The genus-1 simple Hurwitz numbers satisfy
	{\small \barray\label{eq:vh g=1}
	H^1_d &=& 
	\frac{4}{3}\, \sum^{d-1}_{k=1}
\left(\hspace{-1.5mm}\begin{array}{c} 
			2d-1\\ 2k-2 \end{array}\hspace{-1.5mm}
	\right)k\, (d-k) 
	(2k-1) (2d-2k+1) \,H^0_{k} H^1_{d-k}\nonumber\\
	&& \ \  +\,\frac{ d (d-1) (2d-1)^2 }{18} H^0_{d}
.
	\earray}
\end{claim}
It can be easily checked that these relations are actually 
different from 
\eqr{eq:g=0 hurwitz} and \eqr{eq:g=1 hurwitz} and from the ones
obtained by Graber and Pandharipande.  {\bf Remark}: {\it It is important
to note that since the Virasoro conjecture has been proven to hold in
genera-zero and -one, 
the recursion relations \eqr{eq:vh g=0} and \eqr{eq:vh g=1} are also
true and are not mere conjectures.  Indeed, we have verified
numerically that they lead to the correct simple Hurwitz numbers.}
Further investigation is needed to gain a geometric understanding of the
recursion relations that we have obtained.

It is also possible to obtain 
similar relations for higher genus simple Hurwitz numbers from
\eqr{eq:vir-hur}, but there are two important distinctions from the
above two cases.  Firstly, the Virasoro constraints are still
conjectural in genus-2 and higher, thus the resulting recursions are
not rigorous, even though they will provide an interesting check for
the conjecture.  Secondly, there are no effective TRRs that can be
used to express $\ccor{\ds{2,1}\ds{1,1}^{2d+2g-4}}{g}$ and  
$\ccor{\ds{2,0} \ds{1,1}^{2d+2g-3}}{g}$ in terms of Hurwitz numbers.  
In principle, Getzler's 
TRRs \eqr{eq:app genus-2} in genus-2 could be used  to express
these quantities in terms of lower genus simple Hurwitz numbers
and $H^2_k$, for $k<d$.  The TRRs however involve a large number of
terms and render computations somewhat intractable.  
As there already exists a much simpler
recursion relation \cite{GJV}, we omit the derivation here.  In higher 
genus, we are not aware of any effective TRRs that can be applied.  
What seems to be required in this study is a TRR that eliminates the
descendant $\ds{2,\alpha}$ from correlators, just as the string,
divisor and dilaton equations eliminate the $\ds{0,0}$, $\ds{0,1}$ and
$\ds{1,0}$ insertions, respectively.  
It would be interesting to see if there exists
a geometric reason for such an equation.

%%%%%%%%%%%%%%%%%%########################################################
%  Conclusion
%%%%%%%%%%%%%%%%%%########################################################

\section{Conclusion}
In principle, one could, with much patience and stamina, extend 
our program to higher genera.  Being novices that we are in computer
science, we however stop at genus-2 and would now like to discuss what
we have learned from these exercises.

It is instructive to recall how the KdV conjecture 
for pure gravity, stating that the intersection theory of tautological
classes on $\overline{\cM}_{g.n}$ is governed by the KdV hierarchy and
the string equation, was proven by Witten in
genera-zero and -one \cite{Witten}.  First, recall the
algebro-geometric way of determining the descendant integrals:
  In genera-zero and -one, purely dimensional
arguments require the  non-vanishing
descendant integrals to include a certain number of puncture and
dilaton operators.  Then, the string and dilaton equations are used to
reduce the integrals to $\ccor{\ds{0,0}^3}{0}$ in genus-0 and
$\ccor{\ds{1,0}}{1}$ in genus-1, whose values can be determined from
algebraic geometry.  Witten's proof is based on the fact that the string
and dilaton equations and the initial values of
$\ccor{\ds{0,0}^3}{0}$ and $\ccor{\ds{1,0}}{1}$, which together
determine all the descendant integrals completely, can be
derived\footnote{More precisely, the dilaton equation can be
derived from the string and the KdV equations.} from
his KdV conjecture.    
Hence, the algebraic geometry and his KdV conjecture
yield the precisely same algorithm for computing the descendant
integrals in genera-zero and -one. 
In the case of  point target space, there is thus no further need to 
invoke additional topological recursion relations, which are
nevertheless consistent with the KdV structure.

Something similar but crucially different persists in the picture of the
Virasoro constraints 
for $\P^1$ in low genera.  One could compute all the GW-invariants in
genera-zero, -one and -two by using only the string, divisor and dilaton
equations together with the aforementioned topological recursion
relations.  On the other hand, the Virasoro constraints are not
strong enough to determine the GW-invariants by themselves.  It thus
seems that the Virasoro constraints are {\it weaker} \/  than the
topological recursion relations.  As we have checked numerically, 
the GW-invariants obtained from the
TRRs satisfy the Virasoro constraints in genus-zero and -one, as
they should according to the rigorous proofs of mathematicians, and
even in genus-two, which has yet no direct proof.  Thus, as in the
pure gravity case, the TRRs are
consistent with the conjectured integrable hierarchy manifested by the 
Virasoro constraints which however, unlike the pure gravity case, do not
determine the generating functions completely.

The relations between the TRRs and the Virasoro constraints appear to
be quite mysterious.  Even in the case of pure gravity, although
the TRRs seem redundant,  it is not known how to derive them
directly from the KdV hierarchy or whether it is possible to do so at
all.  An analogous question in  the case of $\P^1$, in which the TRRs
and the Virasoro constraints reverse their roles in some sense, would
be: 
{\it Do TRRs imply Virasoro constraints in genera-zero, -one, and
-two?}   Since $L_{-1}$ and
$L_{2}$ generate the (half branch of) Virasoro algebra and since 
$L_{-1}$ is just the string equation, which is true for general
topological string theories, in order to answer the question,
one only needs to prove that the TRRs imply the $L_{2}$ condition.  We
have tried an inductive approach to show that all derivatives of the
$L_{2}Z$ vanish by the TRRs and the $L_{-1}$ constraint, but it does not
seem possible to prove the statement. 

The study of Virasoro constraints is presently only at its rudimentary
stage, and any subsequent effort to understand their hidden structure
would require unraveling their relation with various topological
recursion relations and also with the constraints arising from the
study of Hodge integrals.    
In this paper, we have used the $L_{-1}, L_{0}$ conditions and TRRs to
compute the descendant GW-invariants of $\P^1$ in low genera.  
The ineffectiveness of the Virasoro constraints suggests that there may
exist an enlarged algebra including the
Virasoro algebra and giving us a ``master'' hierarchy encoding 
the TRR relations in all genera. 
Since the 
virtual localization technique expresses  all GW-invariants in
terms of Hodge integrals over products of the moduli space of stable
pointed curves \cite{GrP}, it is also tempting to speculate that the
Virasoro conjecture and the TRRs can be translated into a statement of
some kind of an integrable hierarchy involving the {\it very large}
phase space of Manin and Zograf \cite{MZ}.

\vspace{1cm}
\noindent
{\bf Acknowledgments}

\vspace{1mm}
\noindent
We are grateful to G. Tian and R. Vakil for
numerous valuable discussions and their suggestions.  

We would like to thank Stefano Monni for discussions and for 
his critical
reading of the manuscript throughout this work.
We also thank E. Getzler for a correspondence regarding his
topological recursion relations and Y.S. Song for comments on the
preliminary version of the paper.

%%%%%%%%%%%%%%%%%#########################################
%  Appendix
%%%%%%%%%%%%%%%%%#########################################
\setequation
\newpage
\appendix
\section{Topological Recursion Relations}
We here summarize the  topological recursion relations that form the essential bases for
our discussion.

\vspace{2mm}
\noindent
\underline{{\sc Genus-Zero}} \cite{Witten}:
	\barray
	\cor{\ds{m_1,\alpha_1} \ds{m_2,\alpha_2} \ds{m_3,\alpha_3}}{0}
	&=& \cor{\ds{m_1-1,\alpha_1} \ds{0,0}}{0}\cor{\ds{0,1}
	\ds{m_2,\alpha_2} \ds{m_3,\alpha_3}}{0}\nonumber\\
	&&\hspace{5mm} +\ \cor{\ds{m_1-1,\alpha_1} \ds{0,1}}{0}\cor{\ds{0,0}
	\ds{m_2,\alpha_2} \ds{m_3,\alpha_3}}{0} \ .\label{app-eq:string}
	\earray

\vspace{2mm}
\noindent
\underline{{\sc Genus-One}} \cite{Witten}:
	\beq	\label{eq:app-genus-1TRR}
	\cor{\ds{n,\alpha}}{1} = \cor{\ds{n-1,\alpha} \ds{0,0}}{0}
	\cor{\ds{0,1}}{1} + \cor{\ds{n-1,\alpha} \ds{0,1}}{0}
	\cor{\ds{0,0}}{1} + \oover{12} \cor{\ds{n-1,\alpha} \ds{0,0}
	\ds{0,1}}{0} \ .
	\eeq

\vspace{2mm}
\noindent
\underline{{\sc Genus-Two}} \cite{Getzler2}:
	\barray
	\cor{\ds{k+2,\alpha}}{2} &=&
	\cor{\ds{k+1,\alpha}\ds{0,a}}{0}\eta^{ab}\cor{\ds{0,b}}{2} 
  +  \cor{\ds{k,\alpha} \ds{0,a}}{0}\eta^{ab} \cor{\ds{1,b}}{2}\nonumber\\
  && - \cor{\ds{k,\alpha}\ds{0,a}}{0}\eta^{ab}
	\cor{\ds{0,b}\ds{0,c}}{0} \eta^{cd} \cor{\ds{0,d}}{2} 
	+ \frac{7}{10}\, \cor{\ds{k,\alpha}\ds{0,a}\ds{0,c}}{0} \eta^{ab}
	\cor{\ds{0,b}}{1} \eta^{cd} \cor{\ds{0,d}}{1} \nonumber\\
 &&+ \frac{1}{10}\,
\cor{\ds{k,\alpha}\ds{0,a}\ds{0,c}}{0}\eta^{ab}\eta^{cd}\cor{\ds{0,b}\ds{0,d}}{1} -
\frac{1}{240}\, \cor{\ds{k,\alpha}\ds{0,a}}{1} \eta^{ab}\eta^{cd}
\cor{\ds{0,b}\ds{0,c}\ds{0,d}}{0} \nonumber\\
 	&&+ \frac{13}{240}\,
\cor{\ds{k,\alpha}\ds{0,a}\ds{0,b}\ds{0,c}}{0}\eta^{ab}\eta^{cd} \cor{\ds{0,d}}{1} +
\frac{1}{960}\, \cor{\ds{k,\alpha}\ds{0,a}\ds{0,b}\ds{0,c}\ds{0,d}
}{0}\eta^{ab}\eta^{cd}\ ,\nonumber\\ \label{eq:app genus-2}
\earray
where the metric is given by $\eta_{ab} = \delta_{a,1-b}$.

\section{Genus-Zero Two-Point Descendants} \label{sec:appendix-genus0}
The GW-invariants of the form $\ccor{\ds{m,\alpha}\ds{0,\beta}}{0}$
are easily obtained by using the following TRR for two-point functions
\cite{EHX} which is valid in the large phase space:
	\beq \label{eq:TRR-EHX}
	\cor{\ds{n,\alpha}\ds{0,\beta}}{0} =
\frac{1}{n+\alpha + \beta } \left[ M_{\beta}^{\gamma} 
	  \cor{\ds{n-1,\alpha}\ds{0,\gamma}}{0} -
	2\cor{\ds{n-1,\alpha+1}\ds{0,\beta}}{0}\right]\ ,
	\eeq
where the matrix $M$ is given by
	\beq
	M_{\alpha}^{\beta} = \left(
	\begin{array}{cc}
	\cor{\ds{0,0}\ds{0,1}}{0} & 2 \\
	2 \,\cor{\ds{0,1}\ds{0,1}}{0}& \cor{\ds{0,0}\ds{0,1}}{0}\\
	\end{array}\right) .
	\eeq
When all the couplings are turned off, the matrix $M$
takes the form
	\beq
   M_{\alpha}^{\beta} = \left(
	\begin{array}{cc}
	0 & 2 \\
	2 & 0\\
	\end{array}\right)\ .
	\eeq
From \eqr{eq:TRR-EHX}, one finds
	\barray \label{eq:app g=0 2pt}
	\ccor{\ds{2n+1,1} \ds{0,0}}{0,n+1} &=&
	\ccor{\ds{2n,1}}{0,n+1}\,=\,\left[
		\frac{1}{(n+1)!}\right]^2\nonumber\\ 
	\ccor{\ds{2n,1} \ds{0,1} }{0,n+1} &=&\left(\frac{1}{n+1}\right)
		\frac{1}{(n!)^2} \nonumber\\ 
	\ccor{\ds{2n,0}\ds{0,0} }{0,n} &=& \ccor{\ds{2n-1,0}}{0,n}\,=\,
		-\frac{2\,c_n}{(n!)^2} \nonumber\\
	\ccor{\ds{2n+1,0} \ds{0,1} }{0,n+1} &=& \left(\frac{1}{n+1}\right)
		\frac{1}{(n!)^2}\left[-2\,c_n-
		\frac{1}{n+1}\right]. 
	\earray

For more general invariants, we use the approach discussed in \S\ref{subsec:g=0 cor}:  
	\barray
	\ccor{\ds{2m,0}
\ds{2d-2m,0}}{0,d} &=&  -2\, \frac{c_d}{{{d!}^2}} + \sum^m_{k=1} \Delta_1(k,d)  \nonumber\\
	 \ccor{\ds{2m-1,0} \ds{2d-2m+1,0}}{0,d}&=&
 2\, \frac{c_d}{{{d!}^2}} + \sum^m_{k=2} \widetilde{\Delta_1}(k,d)
	\earray
where
	\barray
	\Delta_1(k,d) &=& A(k,d) - A(d-k+1,d) \nonumber\\ 
	\widetilde{\Delta_1}(k,d) &=& A(d-k+1,d) - A(k-1,d)\ .
	\earray

Similarly, we find
 
	\barray
	\ccor{\ds{2m,1} \ds{2d-2m-2,1}}{0,d} &=&
	\frac{1}{d (d-1)!^2} + \sum^m_{k=1} \Delta_2(k,d)
	\nonumber\\
   \ccor{\ds{2m-1,1} \ds{2d-2m-1,1}}{0,d} &=& 
	\frac{d-1}{d (d-1)!^2}
	 + \sum^m_{k=2} \widetilde{\Delta_2}(k,d)\, ,
	\earray
where
	\barray
	\Delta_2(k,d) &=& M(k,d) - M(d-k,d) \nonumber\\
	\widetilde{\Delta_2}(k,d) &=& M(d-k,d) - M(k-1,d)	\nonumber\\
	M(k,d) &=& \oover{(k)!^2 (d-k)
	(d-k-1)!^2} \ .
	\earray

Finally, we have
	\barray
	\ccor{\ds{2m,0} \ds{2d-2m-1,1}}{0,d} &=& 
	\oover{d!^2} + \sum^m_{k=1} \Delta_3(k,d)	
\nonumber\\
    \ccor{\ds{2m-1,0} \ds{2d-2m,1}}{0,d} &=&
	 -\oover{d!^2}+ \sum^{m}_{k=2} \widetilde{\Delta_3}(k,d) \ ,
	\earray
where
	\barray
	\Delta_3(k,d) &=& W_1(k,d) + W_2(k,d) \nonumber\\
	\widetilde{\Delta_3}(k,d) &=& - W_1(k-1,d)-W_2(k,d) \nonumber\\
	W_1(k,d)&=& -\frac{2 c_{k-1} +1/k}{k (k-1)!^2
	(d-k)!^2}\nonumber\\
	W_2(k,d) &=&  \frac{2\, c_{k-1}}{(k-1)!^2 (d-k+1)(d-k)!^2} \ .
	\earray
Note that the summations are set to zero whenever the lower limit exceeds the upper limit.

%%%%%%%%%%%%
%
%%%%%%%%%%%%

\section{Partial Lists of the GW-Invariants} \label{sec:app-lists}
For those who are interested in the numerical values of the
GW-invariants and for the sake of completeness, we here present, in
fine prints, a
few examples of the non-vanishing invariants.  In most cases, 
we omit the ones that
can be reduced by using the string, dilaton, or divisor equations.

\vspace{5mm}

%%%%%%%%%%%
% genus-0 1-point functions
%%%%%%%%%%%
\subsection{Genus-Zero Descendants} \label{sec:lists g=0}
\vspace{2mm}
\noindent
\underline{{\sc 1-Point Descendants}}

\begin{center}
{\tiny
\begin{minipage}[t]{2in}
\begin{tabular}{|r|c|}\hline
$n$ &  $\ccor{\ds{2n+1,0}}{0,n+1}$ \\ \hline\hline
0 &  $  -2$\\ \hline
1&  $  -{3\over 4}$ \\ \hline
2 &  $ -{{11}\over {108}}$ \\ \hline
3 &  $ -{{25}\over {3456}}$ \\ \hline
4 &  $ -{{137}\over {432000}}$ \\ \hline
5&  $  -{{49}\over {5184000}}$ \\ \hline
6 &  $ -{{121}\over {592704000}}$ \\ \hline
7&  $  -{{761}\over {227598336000}}$ \\ \hline
8&  $  -{{7129}\over {165919186944000}}$ \\ \hline
9 &  $ -{{7381}\over {16591918694400000}}$ \\ \hline
10&  $  -{{83711}\over {22083843782246400000}}$\\ \hline
\end{tabular}
\end{minipage} \ \hspace{5mm} \
\begin{minipage}[t]{2in}
\begin{tabular}{|r|c|}\hline
$n$ &$  \ccor{\ds{2n,1}}{0,n+1}$ \\ \hline\hline
0  &$ 1$ \\ \hline
1  &$ {1\over 4}$ \\ \hline
2 &$  {1\over {36}}$ \\ \hline
3 &$  {1\over {576}}$ \\ \hline
4 &$  {1\over {14400}}$ \\ \hline
5 &$  {1\over {518400}}$ \\ \hline
6 &$  {1\over {25401600}}$ \\ \hline
7 &$  {1\over {1625702400}}$ \\ \hline
8 &$  {1\over {131681894400}}$ \\ \hline
9 &$  {1\over {13168189440000}}$ \\ \hline
10&$   {1\over {1593350922240000}}$\\ \hline
\end{tabular}
\end{minipage}}
\end{center}

%%%%%%%%%%
%  genus-0 2-point functions
%%%%%%%%%%
\vspace{2mm}
\noindent
\underline{{\sc 2-Point Descendants}}

\noindent
\begin{center}
\noindent
{\tiny
\begin{minipage}[t]{1.5in}
{\small $I=\ccor{\ds{2n,0}\ds{2d-2n,0}}{0,d}$}
\begin{tabular}{|r|r|r|r|}\hline
$d$ &$ 2n $ & $ 2d-2n $ & $I$ \\ \hline\hline
1 & 0  & 2   & $	 -2 $ \\ \hline
2  &0  &4 	 & $  -{3\over 4} $ \\ \hline
2  &2 & 2   & $ 	{5\over 4} $ \\ \hline
3 & 0 & 6   & $ 	-{{11}\over {108}} $ \\ \hline
3  &2 & 4   & $ 	{{35}\over {54}} $ \\ \hline
4 & 0 & 8  & $ 	 -{{25}\over {3456}} $ \\ \hline
4 & 2 & 6  & $ 	 {{109}\over {1152}} $ \\ \hline
4 & 4  &4  & $ 	 {{61}\over {128}} $ \\ \hline
5 & 0 & 10  & $ 	 -{{137}\over {432000}} $ \\ \hline
5 & 2 & 8   & $	 {{83}\over {12000}} $ \\ \hline
5 & 4 & 6   & $	 {{479}\over {6000}} $ \\ \hline
6 & 0 	&12  & $    -{{49}\over {5184000}} $ \\ \hline
6 & 2 & 10  & $    {{319}\over {1036800}} $ \\ \hline
6 & 4  &8  & $     {{131}\over {20736}} $ \\ \hline
6 & 6  &6  & $     {{221}\over {15552}} $ \\ \hline
\end{tabular}
\end{minipage} \ \hspace{2mm} \
\begin{minipage}[t]{1.6in}
{\small $I=\ccor{\ds{2n-1,0}\ds{2d-2n+1,0}}{0,d}$}
\begin{tabular}{|r|r|r|r|}\hline
$d$ &$2n-1$ &$ 2d-2n+1$	&   	$I$\\ \hline\hline
1 & 1 & 	1  &  	$  2$\\ \hline
2 & 1 & 	3  &  	$ {3\over 4}$\\ \hline
3  &1  &	5  & 	$ {{11}\over {108}}$\\ \hline
3 & 3  &	3 &  	$ {{50}\over {27}}$\\ \hline
4 & 1 & 	7 &  	$ {{25}\over {3456}}$\\ \hline
4 & 3  &	5 &  	$ {{59}\over {128}}$\\ \hline
5 & 1 & 	9  & 	$  {{137}\over {432000}}$\\ \hline
5 & 3  &	7 &  	$  {{641}\over {13500}}$\\ \hline
5 & 5  &	5  & 	$  {{257}\over {2000}}$\\ \hline
6 & 1  &	11  &  $	{{49}\over {5184000}}$\\ \hline
6 & 3	 & 9   & 	${{113}\over {41472}}$\\ \hline
6 & 5  	 &7 &	$  {{73}\over {5184}}$\\ \hline
\end{tabular}
\end{minipage} \ \hspace{6mm} \
\begin{minipage}[t]{1.5in}
{\small $I= \ccor{\ds{2n,1}\ds{2d-2n-2,1}}{0,d}$}
\begin{tabular}{|r|r|r|r|}\hline
$d$ & $2n$ &  $ 2d-2n-2$ &  $I$\\
\hline \hline
1 & 0 	 &0   &	$ 1$\\ \hline
2 & 0  &	2   &	$ {1\over 2}$\\ \hline
3 & 0  &	4  &  	${1\over {12}}$\\ \hline
3 & 2  &	2  & 	$ {1\over 3}$\\ \hline
4 & 0  &	6  & 	$ {1\over {144}}$\\ \hline
4 & 2  &	4  &  	${1\over {16}}$\\ \hline
5 & 0 & 	8  & 	$ {1\over {2880}}$\\ \hline
5 & 2 & 	6  & 	$ {1\over {180}}$\\ \hline
5 & 4  &	4  & 	$ {1\over {80}}$\\ \hline
6 & 0  &    10   & 	${1\over {86400}}$\\ \hline
6 & 2  &	8   &  $	{1\over {3456}}$\\ \hline
6 & 4  &	6   &  $	{1\over {864}}$\\ \hline
\end{tabular}
\end{minipage}}
\end{center}

\vspace{5mm}
\begin{center}
{\tiny
\begin{minipage}[t]{1.6in}
{\small $I=\ccor{\ds{2n-1,1}\ds{2d-2n-1,1}}{0,d}$}
\begin{tabular}{|r|r|r|r|}\hline
$d$ & $2n-1$  &  $2d-2n-1$ & $I$\\ \hline\hline
2 & 1 	 &	1  & $ {1\over 2}$\\ \hline
3 & 1  &		3 & $  {1\over 6}$\\ \hline
4 & 1  &		5  & $ {1\over {48}}$\\ \hline
4 & 3  &		3  & $ {1\over {16}}$\\ \hline
5 & 1  &		7  & $ {1\over {720}}$\\ \hline
5 & 3  &		5  & $ {1\over {120}}$\\ \hline
6 & 1  &		9  & $ {1\over {17280}}$\\ \hline
6 & 3  &		7  & $ {1\over {1728}}$\\ \hline
6 & 5  &		5  & $ {1\over {864}}$\\ \hline
\end{tabular}
\end{minipage} \ \hspace{5mm}
\begin{minipage}[t]{1.5in}
{\small $I =\ccor{\ds{2n,0}\ds{2d-2n-1,1}}{0,d}$}
\begin{tabular}{|r|r|r|r|}\hline
$d$ &$2n$ &   $2d-2n-1$   & $I$\\
\hline \hline
1& 0 	&	1 & $ 	 1$\\ \hline
2& 0 	&	3 & $      {1\over 4}$\\ \hline
2 &2 	&	1 & $   -{3\over 4}$\\ \hline
3& 0	&	5 & $  {1\over {36}}$\\ \hline
3& 2 	&	3 & $  -{2\over 9}$\\ \hline
3& 4 	&	1 & $  -{{17}\over {36}}$\\ \hline
4& 0 	&	7 & $   {1\over {576}}$\\ \hline
4& 2 	&	5 & $   -{5\over {192}}$\\ \hline
4& 4 	&	3 & $   -{{11}\over {64}}$\\ \hline
4& 6 	&	1 & $   -{{43}\over {576}}$\\ \hline
5& 0 	&9   & $ {1\over {14400}}$\\ \hline
5& 2 	&	7& $    -{1\over {600}}$\\ \hline
5& 4 	&	5 & $   -{9\over {400}}$\\ \hline
5& 6 	&	3 & $   -{{53}\over {1800}}$\\ \hline
5& 8 	&	1 & $   -{{247}\over {43200}}$\\ \hline
\end{tabular}
\end{minipage} \ \hspace{5mm} \
\begin{minipage}[t]{1.5in}
{\small $I=\ccor{\ds{2n-1,0}\ds{2d-2n,1}}{0,d}$}
\begin{tabular}{|r|r|r|r|}\hline
$d$ & $2n-1$ &   $2d-2n$  &  $I$\\
\hline \hline
1& 1& 0&   $   -1$\\ \hline
2& 1& 2 &   $  -{1\over 4}$\\ \hline
2& 3& 0&   $   -{5\over 4}$\\ \hline
3& 1& 4&   $   -{1\over {36}}$\\ \hline
3& 3& 2 &   $  -{7\over 9}$\\ \hline
3& 5& 0&   $   -{5\over {18}}$\\ \hline
4& 1& 6 &   $  -{1\over {576}}$\\ \hline
4& 3& 4&   $   -{9\over {64}}$\\ \hline
4& 5& 2&   $   -{{13}\over {64}}$\\ \hline
4& 7& 0&   $   -{{47}\over {1728}}$\\ \hline
5& 1& 8&   $   -{1\over {14400}}$\\ \hline
5& 3& 6&   $   -{{11}\over {900}}$\\ \hline
5& 5& 4&   $   -{1\over {25}}$\\ \hline
5& 7& 2&   $   -{{29}\over {1350}}$\\ \hline
5& 9& 0 &   $  -{{131}\over {86400}}$\\ \hline
\end{tabular}
\end{minipage}}
\end{center}

%%%%%%%%%
%    g=0 Three Point Functions 
%%%%%%%%%

\vspace{4mm}
\noindent
\underline{{\sc 3-Point Descendants}}
\vspace{2mm}

\noindent
{\tiny
\begin{minipage}[t]{1.2in}
{\small $I=\ccor{\ds{m,0}\ds{n,0}\ds{\ell,0}}{0}$}\vspace{1mm}
\begin{tabular}{|r|r|r|r|}\hline
$m$ & $n$ & $\ell$  &  $I$\\ \hline \hline
2 & 2 & 3 & $  -2$\\ \hline
2 & 2&  5&  $  -{3\over 4}$\\ \hline
2 & 3&  4 & $  -{5\over 2}$\\ \hline
2 & 4&  5 & $  -{{15}\over {16}}$\\ \hline
3 & 3 & 3 & $  -8$\\ \hline
3&  3 & 5 & $  -3$\\ \hline
3&  4 & 4 & $  -{{25}\over 8}$\\ \hline
3 & 5 & 5 & $  -{9\over 8}$\\ \hline
4 & 4&  5 & $  -{{75}\over {64}}$\\ \hline
5 & 5&  5 &  $ -{{27}\over {64}}$\\ \hline
\end{tabular}
\end{minipage} \ \hspace{.5cm} \
\begin{minipage}[t]{1.2in}
{\small $I=\ccor{\ds{m,1}\ds{n,1}\ds{\ell,1}}{0}$}\vspace{1mm}
\begin{tabular}{|r|r|r|r|}\hline
$m$ & $n$&  $\ell $ &  $I$\\ \hline \hline
1 &  1  & 2  & $  1$\\ \hline
1  & 1 &  4  & $  {1\over 4}$\\ \hline
1  & 2 &  3  & $  {1\over 2}$\\ \hline
1 &  2 &  5  & $  {1\over {12}}$\\ \hline
1  & 3 &  4  &  $ {1\over 8}$\\ \hline
1 &  4 &  5  & $  {1\over {48}}$\\ \hline
2 &  2 &  2   & $ 1$\\ \hline
2 &  2 &  4  &  $ {1\over 4}$\\ \hline
2 &  3  & 3  &  $ {1\over 4}$\\ \hline
2 &  3 &  5  &  $ {1\over {24}}$\\ \hline
2 &  4 &  4  &  $ {1\over {16}}$\\ \hline
2 &  5 &  5  &  $ {1\over {144}}$\\ \hline
3  & 3 &  4  &  $ {1\over {16}}$\\ \hline
3  & 4 &  5   & $ {1\over {96}}$\\ \hline
4 &  4 &  4  &  $ {1\over {64}}$\\ \hline
4 &  5 &  5  & $  {1\over {576}}$\\ \hline
5 &  5  & 4  &  $ {1\over {576}}$\\ \hline
\end{tabular}
\end{minipage} \ \hspace{.5cm} \
\begin{minipage}[t]{1.2in}
{\small $I= \ccor{\ds{m,1}\ds{n,0}\ds{\ell,0}}{0}$}\vspace{1mm}
\begin{tabular}{|r|r|r|r|}\hline
$m$ & $n $& $\ell$ &  $ I$\\ \hline \hline
1 & 2&  3 &  $   2$\\ \hline
1 & 2&  5 &  $   {3\over 4}$\\ \hline
1 & 3&  4 &  $   {5\over 2}$\\ \hline
1&  4&  5 &  $   {{15}\over {16}}$\\ \hline
2&  2&  2 &  $   1$\\ \hline
2 & 2&  4 &  $   {5\over 4}$\\ \hline
2&  3&  3 &  $   4$\\ \hline
2&  3&  5 &  $   {3\over 2}$\\ \hline
2&  4&  4 &  $   {{25}\over {16}}$\\ \hline
2&  5 & 5 &  $   {9\over {16}}$\\ \hline
3&  2 & 3 &  $   1$\\ \hline
3&  2&  5 &  $   {3\over 8}$\\ \hline
3 & 3&  4 &  $   {5\over 4}$\\ \hline
3&  4 & 5 &  $   {{15}\over {32}}$\\ \hline
4 & 2&  2 &  $   {1\over 4}$\\ \hline
4&  2&  4 &  $   {5\over {16}}$\\ \hline
4&  3 & 3 &  $   1$\\ \hline
4 & 3&  5&  $    {3\over 8}$\\ \hline
4&  4 & 4&  $    {{25}\over {64}}$\\ \hline
4&  5&  5 &  $   {9\over {64}}$\\ \hline
5&  2&  3 &  $   {1\over 6}$\\ \hline
5&  2 & 5 &  $   {1\over {16}}$\\ \hline
5&  3 & 4&  $    {5\over {24}}$\\ \hline
5 & 4&  5 &  $   {5\over {64}}$\\ \hline
\end{tabular}
\end{minipage} \ \hspace{.5cm} \
\begin{minipage}[t]{1.2in}
{\small $I= \ccor{\ds{m,1}\ds{n,1}\ds{\ell,0}}{0}$}\vspace{1mm}
\begin{tabular}{|r|r|r|r|}\hline
$m$ & $n $ &$\ell$ &  $ I$\\ \hline
\hline
1  & 1 &  3 &  $   -2$\\ \hline
1  & 1  & 5  &  $  -{3\over 4}$\\ \hline
1 &  2 &  2 &  $   -1$\\ \hline
1  & 2 &  4 &  $   -{5\over 4}$\\ \hline
1  & 3 &  3 &  $   -1$\\ \hline
1  & 3 &  5 &  $   -{3\over 8}$\\ \hline
1  & 4 &  2 &  $   -{1\over 4}$\\ \hline
1 &  4 &  4 &  $   -{5\over {16}}$\\ \hline
1  & 5 &  3 &  $   -{1\over 6}$\\ \hline
1  & 5 &  5 &  $   -{1\over {16}}$\\ \hline
2  & 2 &  3 &  $   -2$\\ \hline
2 &  2 &  5 &  $   -{3\over 4}$\\ \hline
2  & 3 &  2 &  $   -{1\over 2}$\\ \hline
2 &  3 &  4 &  $   -{5\over 8}$\\ \hline
2 &  4  & 3&  $    -{1\over 2}$\\ \hline
2  & 4 &  5&  $    -{3\over {16}}$\\ \hline
2  & 5 &  2&  $    -{1\over {12}}$\\ \hline
2  & 5 &  4 &  $   -{5\over {48}}$\\ \hline
3  & 3 &  3 &  $   -{1\over 2}$\\ \hline
3 &  3 &  5 &  $   -{3\over {16}}$\\ \hline
3 &  4 &  2 &  $   -{1\over 8}$\\ \hline
3 &  4 &  4 &  $   -{5\over {32}}$\\ \hline
3 &  5 &  3 &  $   -{1\over {12}}$\\ \hline
3  & 5 &  5 &  $   -{1\over {32}}$\\ \hline
4 &  4 &  3 &  $   -{1\over 8}$\\ \hline
4  & 4 &  5&  $    -{3\over {64}}$\\ \hline
4 &  5 &  2&  $    -{1\over {48}}$\\ \hline
4 &  5 &  4 &  $   -{5\over {192}}$\\ \hline
5 &  5 &  3 &  $   -{1\over {72}}$\\ \hline
5  & 5 &  5 &  $   -{1\over {192}}$\\ \hline
\end{tabular}
\end{minipage}}

%---------

%%%%%%%%%%%%%%%%%%%
%    Genus One
%%%%%%%%%%%%%%%%%%%
\newpage
\subsection{Genus-One Descendants} \label{sec:app-genus1} 

\vspace{2mm}
\noindent
\underline{{\sc 1-Point Descendants}}
\begin{center}
{\tiny
\begin{minipage}[t]{2in}
\begin{tabular}{|r|c|}\hline
$n$ &  $\ccor{\ds{n,0}}{1}$\\ \hline \hline
1  & $  {1\over {12}}$\\ \hline
5   &$ -{7\over {96}}$\\ \hline
7  & $ -{{49}\over {2592}}$\\ \hline
9   & $-{{163}\over {82944}}$\\ \hline
11& $   -{{391}\over {3456000}}$\\ \hline
13 & $  -{{173}\over {41472000}}$\\ \hline
15 & $  -{{4579}\over {42674688000}}$\\ \hline
17& $   -{{2227}\over {1092472012800}}$\\ \hline
19& $   -{{118673}\over {3982060486656000}}$\\ \hline
21& $   -{{137719}\over {398206048665600000}}$\\ \hline
\end{tabular}
\end{minipage} \ \hspace{5mm} \
\begin{minipage}[t]{3in}
\begin{tabular}{|r|c|}\hline
$n$ &  $   \ccor{\ds{n,1}}{1}$\\ \hline \hline
0& $   -{1\over {24}}$\\ \hline
2& $   {1\over {24}}$\\ \hline
4& $   {1\over {32}}$\\ \hline
6& $   {5\over {864}}$\\ \hline
8 & $  {7\over {13824}}$\\ \hline
10& $   {1\over {38400}}$\\ \hline
12& $   {{11}\over {12441600}}$\\ \hline
14& $   {{13}\over {609638400}}$\\ \hline
16& $   {1\over {2601123840}}$\\ \hline
18& $   {{17}\over {3160365465600}}$\\ \hline
20& $   {{19}\over {316036546560000}}$\\ \hline
\end{tabular}
\end{minipage}}
\end{center}

%%%%%%%%
%  g=1 1-point Hodge integral
%%%%%%%%

\vspace{2mm}
\noindent
\underline{{\sc 1-Point Hodge Integrals}}
\begin{center}
{\tiny
\begin{minipage}[t]{2in}
\begin{tabular}{|r|c|}\hline
$m$ &  $\ccor{\ds{m,0}\lambda_1}{1}$\\ \hline \hline
0 &$ {1\over {12}}$\\ \hline
2 &$ -{1\over {12}}$\\ \hline
4 &$ -{5\over {48}}$\\ \hline
6 &$ -{5\over {216}}$\\ \hline
8 &$ -{{47}\over {20736}}$\\ \hline
10&$  -{{131}\over {1036800}}$\\ \hline
12 &$ -{{71}\over {15552000}}$\\ \hline
14 &$ -{{353}\over {3048192000}}$\\ \hline
16 &$ -{{1487}\over {682795008000}}$\\ \hline
18 &$ -{{6989}\over {221225582592000}}$\\ \hline
20 &$ -{{1451}\over {3982060486656000}}$\\ \hline
\end{tabular}
\end{minipage} \ \hspace{5mm} \
\begin{minipage}[t]{3in}
\begin{tabular}{|r|c|}\hline
$m$ &$ \ccor{\ds{m,1}\lambda_1}{1}$\\ \hline \hline
1 &$ {1\over {12}}$\\ \hline
3 &$ {1\over {24}}$\\ \hline
5 &$ {1\over {144}}$\\ \hline
7 &$ {1\over {1728}}$\\ \hline
9 &$ {1\over {34560}}$\\ \hline
11&$  {1\over {1036800}}$\\ \hline
13 &$ {1\over {43545600}}$\\ \hline
15 &$ {1\over {2438553600}}$\\ \hline
17 &$ {1\over {175575859200}}$\\ \hline
19 &$ {1\over {15801827328000}}$\\ \hline
21 &$ {1\over {1738201006080000}}$\\ \hline
\end{tabular}
\end{minipage}}
\end{center}

%%%%%%%%%%%
%   g=1 Two Point Functions
%%%%%%%%%%%

\vspace{2mm}
\noindent
\underline{{\sc 2-Point Descendants}}

\noindent
\begin{center}
{\tiny
\begin{minipage}[t]{1.5in}
\begin{tabular}{|r|r|r|}\hline
$m$ & $ n$ &  $ \ccor{\ds{m,0}\ds{n,0}}{1}$\\ \hline \hline
1 & 1 & $ {1\over {12}}$\\ \hline
1 & 5 & $ -{7\over {96}}$\\ \hline
1 & 7 & $ -{{49}\over {2592}}$\\ \hline
1 & 9 & $ -{{163}\over {82944}}$\\ \hline
2 & 2 & $ -{1\over 6}$\\ \hline
2 & 4 & $ -{5\over {48}}$\\ \hline
2 & 6 &$  {{43}\over {864}}$\\ \hline
2 & 8 &$  {{115}\over {6912}}$\\ \hline
3 & 3 &$  -{1\over {12}}$\\ \hline
3 & 5 &$  {{29}\over {96}}$\\ \hline
3 & 7 &$  {{313}\over {2592}}$\\ \hline
3 & 9 &$  {{461}\over {27648}}$\\ \hline
4 & 4 &$  {1\over 3}$\\ \hline
4 & 6 & $ {{83}\over {384}}$\\ \hline
4 & 8 & $ {{421}\over {10368}}$\\ \hline
5 & 5 & $ {{29}\over {96}}$\\ \hline
5 & 7 & $ {{817}\over {10368}}$\\ \hline
5 & 9 & $ {{1015}\over {110592}}$\\ \hline
6 & 6 & $ {{151}\over {1728}}$\\ \hline
6 & 8 &$  {{20305}\over {1492992}}$\\ \hline
7 & 7 &$  {{49}\over {2916}}$\\ \hline
7 & 9 &$  {{7825}\over {4478976}}$\\ \hline
8 & 8 &$  {{157}\over {82944}}$\\ \hline
\end{tabular}
\end{minipage} \ \hspace{5mm} \
\begin{minipage}[t]{1.5in}
\begin{tabular}{|r|r|r|}\hline
$m$ &  $n $ &  $\ccor{\ds{m,1}\ds{n,0}}{1}$\\ \hline\hline
1 & 2 & $ {1\over 8}$\\ \hline
1 & 4 &$  -{{11}\over {96}}$\\ \hline
1 & 6 &$  -{{107}\over {864}}$\\ \hline
1 & 8 &$  -{{121}\over {4608}}$\\ \hline
2 & 1 &$  {1\over {24}}$\\ \hline
2 & 3 &$  -{1\over 8}$\\ \hline
2 & 5 &$  -{1\over 4}$\\ \hline
2 & 7 &$  -{{203}\over {2592}}$\\ \hline
2 & 9 &$  -{{823}\over {82944}}$\\ \hline
3 & 4 &$  -{7\over {32}}$\\ \hline
3 & 6 &$  -{{191}\over {1728}}$\\ \hline
3 & 8 & $ -{{791}\over {41472}}$\\ \hline
4 & 1 &$  {1\over {32}}$\\ \hline
4 & 3 &$  -{5\over {32}}$\\ \hline
4 & 5 &$  -{{49}\over {384}}$\\ \hline
4 & 7 &$  -{{53}\over {1728}}$\\ \hline
4 & 9 &$  -{{71}\over {20736}}$\\ \hline
5 & 2 & $ -{7\over {288}}$\\ \hline
5 & 4 &$  -{{31}\over {384}}$\\ \hline
5 & 6 &$  -{{35}\over {1152}}$\\ \hline
5 & 8 & $ -{{2263}\over {497664}}$\\ \hline
6 & 1 &$  {5\over {864}}$\\ \hline
6 & 3 &$  -{{35}\over {864}}$\\ \hline
6 & 5 &$  -{{43}\over {1728}}$\\ \hline
6 & 7 &$  -{{53}\over {10368}}$\\ \hline
6 & 9 & $ -{{1549}\over {2985984}}$\\ \hline
7 & 2 &$  -{1\over {192}}$\\ \hline
7 & 4 &$  -{{41}\over {3456}}$\\ \hline
7 & 6 &$  -{{479}\over {124416}}$\\ \hline
7 & 8 &$  -{{29}\over {55296}}$\\ \hline
8 & 1 &$  {7\over {13824}}$\\ \hline
8 & 3 &$  -{7\over {1536}}$\\ \hline
8 & 5 &$  -{5\over {2048}}$\\ \hline
8 & 7 &$  -{{85}\over {186624}}$\\ \hline
\end{tabular}
\end{minipage} \ \hspace{5mm} \
\begin{minipage}[t]{1.5in}
\begin{tabular}{|r|r|r|}\hline
$m$ &  $n$ &   $\ccor{\ds{m,1}\ds{n,1}}{1}$\\ \hline\hline
1 & 3 &$  {5\over {48}}$\\ \hline
1 & 5 & $ {7\over {144}}$\\  \hline
1 & 7 &$  {1\over {128}}$\\ \hline
1 & 9 &$  {{11}\over {17280}}$\\ \hline
2 & 2 &$  {1\over 6}$\\ \hline
2 & 4 &$  {{11}\over {96}}$\\ \hline
2 & 6 &$  {{11}\over {432}}$\\ \hline
2 & 8 &$  {{37}\over {13824}}$\\ \hline
3 & 3 &$  {1\over 8}$\\ \hline
3 & 5 &$  {{23}\over {576}}$\\ \hline
3 & 7 & $ {{19}\over {3456}}$\\ \hline
3 & 9 &$  {{19}\over {46080}}$\\ \hline
4 & 4 &$  {5\over {96}}$\\ \hline
4 & 6 &$  {{11}\over {1152}}$\\ \hline
4 & 8 & $ {{25}\over {27648}}$\\ \hline
4 & 10 &$  {{71}\over {1382400}}$\\ \hline
5 & 5 &$  {1\over {96}}$\\ \hline
5 & 7 &$  {{53}\over {41472}}$\\ \hline
5 & 9 &$  {{37}\over {414720}}$\\ \hline
6 & 2 &$  {{11}\over {432}}$\\ \hline
6 & 4 & $ {{11}\over {1152}}$\\ \hline
6 & 6 & $ {1\over {648}}$\\ \hline
6 & 8 &$  {{67}\over {497664}}$\\ \hline
7 & 7 &$  {1\over {6912}}$\\ \hline
7 & 9 &$  {{19}\over {1990656}}$\\ \hline
8 & 8 &$  {{11}\over {995328}}$\\ \hline
\end{tabular}
\end{minipage}}
\end{center}

%%%%%%%%%%
%  g=1 2-point Hodge integrals
%%%%%%%%%%

\vspace{2mm}
\noindent
\underline{{\sc 2-Point Hodge Integrals}}

\noindent
\begin{center}
{\tiny
\begin{minipage}[t]{1.5in}
\begin{tabular}{|r|r|r|}\hline
$m$ & $ n$ &  $ \ccor{\ds{m,0}\ds{n,0}\lambda_1}{1}$\\ \hline \hline
0 &  1&  $   {1\over {12}}$\\ \hline
0 & 3 &  $  -{1\over {12}}$\\ \hline
0 & 5&  $   -{5\over {48}}$\\ \hline
0 & 7&  $   -{5\over {216}}$\\ \hline
0 & 9&  $   -{{47}\over {20736}}$\\ \hline
1 & 2&  $   -{1\over {12}}$\\ \hline
1 & 4&  $   -{5\over {48}}$\\ \hline
1 & 6 &  $  -{5\over {216}}$\\ \hline
1 & 8 &  $  -{{47}\over {20736}}$\\ \hline
1 & 10&  $   -{{131}\over {1036800}}$\\ \hline
2 & 3 &  $  {1\over {12}}$\\ \hline
2 & 5&  $   {5\over {48}}$\\ \hline
2 & 7&  $   {5\over {216}}$\\ \hline
2 & 9&  $   {{47}\over {20736}}$\\ \hline
3 & 4&  $   {7\over {16}}$\\ \hline
3 & 6&  $   {4\over {27}}$\\ \hline
3 & 8&  $   {{133}\over {6912}}$\\ \hline
3 & 10&  $   {{1381}\over {1036800}}$\\ \hline
4 & 5&  $   {{49}\over {192}}$\\ \hline
4 & 7&  $   {{119}\over {2592}}$\\ \hline
4 & 9&  $   {{335}\over {82944}}$\\ \hline
5 & 6&  $   {{131}\over {1728}}$\\ \hline
5 & 8&  $   {{763}\over {82944}}$\\ \hline
5 & 10&  $   {{253}\over {414720}}$\\ \hline
6 & 7&  $   {{199}\over {15552}}$\\ \hline
6 & 9&  $   {{1615}\over {1492992}}$\\ \hline
7 & 8&  $   {{1673}\over {1119744}}$\\ \hline
7 & 10&  $   {{2161}\over {22394880}}$\\ \hline
8 & 9&  $   {{4409}\over {35831808}}$\\ \hline
9 & 10&  $   {{9313}\over {1194393600}}$\\ \hline
\end{tabular}
\end{minipage} \ \hspace{5mm} \
\begin{minipage}[t]{1.5in}
\begin{tabular}{|r|r|r|}\hline
$m$ &  $n $ &  $\ccor{\ds{m,1}\ds{n,1}\lambda_1}{1}$\\ \hline\hline
0 & 1 &  $  {1\over {12}}$\\ \hline
0 & 3 &  $  {1\over {12}}$\\ \hline
0 & 5&  $   {1\over {48}}$\\ \hline
0 & 7&  $   {1\over {432}}$\\ \hline
0 & 9&  $   {1\over {6912}}$\\ \hline
1 & 2&  $   {1\over 6}$\\ \hline
1 & 4&  $   {1\over {16}}$\\ \hline
1 & 6&  $   {1\over {108}}$\\ \hline
1 & 8&  $   {5\over {6912}}$\\ \hline
1 & 10&  $   {1\over {28800}}$\\ \hline
2 & 3&  $   {1\over 8}$\\ \hline
2 & 5 &  $  {1\over {36}}$\\ \hline
2 & 7 &  $  {5\over {1728}}$\\ \hline
2 & 9 &  $  {1\over {5760}}$\\ \hline
3 & 4 &  $  {1\over {24}}$\\ \hline
3 & 6 &  $  {5\over {864}}$\\ \hline
3 & 8&  $   {1\over {2304}}$\\ \hline
3 & 10 &  $  {7\over {345600}}$\\ \hline
4 & 5 &  $  {5\over {576}}$\\ \hline
4 & 7&  $   {1\over {1152}}$\\ \hline
4 & 9 &  $  {7\over {138240}}$\\ \hline
5 & 6 &  $  {1\over {864}}$\\ \hline
5 & 8 &  $  {7\over {82944}}$\\ \hline
5 & 10 &  $  {1\over {259200}}$\\ \hline
6 & 7 &  $  {7\over {62208}}$\\ \hline
6 & 9&  $   {1\over {155520}}$\\ \hline
7 & 8&  $   {1\over {124416}}$\\ \hline
7 & 10 &  $  {1\over {2764800}}$\\ \hline
8 & 9 &  $  {1\over {2211840}}$\\ \hline
9 & 10 &  $  {1\over {49766400}}$\\ \hline
\end{tabular}
\end{minipage} \ \hspace{5mm} \
\begin{minipage}[t]{1.5in}
\begin{tabular}{|r|r|r|}\hline
$m$ &  $n$ &   $\ccor{\ds{m,0}\ds{n,1}\lambda_1}{1}$\\ \hline\hline
0& 2 &   $  {1\over {12}}$\\ \hline
0& 4 &   $  {1\over {24}}$\\ \hline
0& 6 &   $  {1\over {144}}$\\ \hline
1& 1 &   $  {1\over {12}}$\\ \hline
1& 3 &   $  {1\over {24}}$\\ \hline
1& 5 &   $  {1\over {144}}$\\ \hline
2& 2 &   $  -{1\over {12}}$\\ \hline
2& 4&   $   -{1\over {24}}$\\ \hline
2& 6&   $   -{1\over {144}}$\\ \hline
3& 1 &   $  -{1\over 4}$\\ \hline
3& 3&   $   -{5\over {24}}$\\ \hline
3& 5&   $   -{7\over {144}}$\\ \hline
4& 0 &   $  -{1\over 6}$\\ \hline
4& 2 &   $  -{{13}\over {48}}$\\ \hline
4 &4&   $   -{3\over {32}}$\\ \hline
4& 6 &   $  -{{23}\over {1728}}$\\ \hline
5& 1 &   $  -{1\over 6}$\\ \hline
5& 3&   $   -{{11}\over {96}}$\\ \hline
5& 5&   $   -{7\over {288}}$\\ \hline
6& 0&   $   -{1\over {16}}$\\ \hline
6& 2&   $   -{{37}\over {432}}$\\ \hline
6& 4&   $   -{{47}\over {1728}}$\\ \hline
6& 6&   $   -{{19}\over {5184}}$\\ \hline
\end{tabular}
\end{minipage}}
\end{center}

%%%%%%%%%%%%%%%%
%    Three Point Functions
%%%%%%%%%%%%%%%%
\newpage
\vspace{2mm}
\noindent
\underline{{\sc 3-Point Descendants}}
\vspace{2mm}

\noindent
{\tiny
\begin{minipage}[t]{1.2in}
{\small $I=\ccor{\ds{m,0}\ds{n,0}\ds{\ell,0}}{1}$}\vspace{1mm}
\begin{tabular}{|r|r|r|r|}\hline
$m$ & $ n$ & $ \ell$ &$ I$\\ \hline\hline
2 & 2 & 3 & $  {1\over 4}$\\ \hline
2 & 2 & 5 & $  {{17}\over {48}}$\\ \hline
2 & 3 & 4 & $  {{23}\over {24}}$\\ \hline
2 & 3 & 6 &$   -{{17}\over {216}}$\\ \hline
2 & 4 & 5 & $  -{{21}\over {128}}$\\ \hline
2 & 5 & 6 & $  -{{1297}\over {3456}}$\\ \hline
3 & 3 & 3 & $  {4\over 3}$\\ \hline
3 & 3 & 5 & $  -{{71}\over {48}}$\\ \hline
3 & 4 & 4 & $  -{{319}\over {192}}$\\ \hline
3 & 4 & 6 & $  -{7\over 4}$\\ \hline
3 & 5 & 5 & $  -{{473}\over {192}}$\\ \hline
3 & 6 & 6 &$   -{{7159}\over {7776}}$\\ \hline
4 & 4 & 5 & $  -{{187}\over {64}}$\\ \hline
4 & 5 & 6 & $  -{{21331}\over {13824}}$\\ \hline
5 & 5 & 5 & $  -{{1681}\over {768}}$\\ \hline
5 & 6 & 6 & $  -{{1105}\over {1728}}$\\ \hline
\end{tabular}
\end{minipage} \ \hspace{.5cm} \
\begin{minipage}[t]{1.2in}
{\small $I=\ccor{\ds{m,1}\ds{n,1}\ds{\ell,1}}{1}$}\vspace{1mm}
\begin{tabular}{|r|r|r|r|}\hline
$m$ & $n$&  $\ell $ &  $I$\\ \hline \hline
1 & 1 & 2 & $  {7\over {24}}$\\ \hline
1 & 1 & 4 & $  {3\over 8}$\\ \hline
1 & 1 & 6 &  $ {{11}\over {96}}$\\ \hline
1 & 2 & 3 & $  {{31}\over {48}}$\\ \hline
1 & 2 & 5 &  $ {{85}\over {288}}$\\ \hline
1 & 3 & 4 &  $ {{41}\over {96}}$\\ \hline
1 & 3 & 6 &  $ {{175}\over {1728}}$\\ \hline
1 & 4 & 5 &  $ {{83}\over {576}}$\\ \hline
1 & 5 & 6 &  $ {{11}\over {384}}$\\ \hline
2 & 2 & 2 &  $ {{25}\over {24}}$\\ \hline
2 & 2 & 4 &  $ {{17}\over {24}}$\\ \hline
2 & 2 & 6 &  $ {{149}\over {864}}$\\ \hline
2 & 3 & 3 &  $ {{25}\over {32}}$\\ \hline
2 & 3 & 5 &  $ {{157}\over {576}}$\\ \hline
2 & 4 & 4 &  $ {{137}\over {384}}$\\ \hline
2 & 4 & 6 & $  {7\over {96}}$\\ \hline
2 & 5 & 5 &   ${{275}\over {3456}}$\\ \hline
2 & 6 & 6 &   ${{409}\over {31104}}$\\ \hline
3 & 3 & 4 &   ${{19}\over {48}}$\\ \hline
3 & 3 & 6 &   ${{31}\over {384}}$\\ \hline
3 & 4 & 5 &   ${{67}\over {576}}$\\ \hline
3 & 5 & 6 &   ${{437}\over {20736}}$\\ \hline
4 & 4 & 4 &   ${{59}\over {384}}$\\ \hline
4 & 4 & 6 &   ${{389}\over {13824}}$\\ \hline
4 & 5 & 5 &   ${{53}\over {1728}}$\\ \hline
4 & 6 & 6 &   ${{49}\over {10368}}$\\ \hline
5 & 5 & 6 &   ${{71}\over {13824}}$\\ \hline
5 & 6 & 5 &   ${{71}\over {13824}}$\\ \hline
6 & 6 & 6 &   ${{31}\over {41472}}$\\ \hline
\end{tabular}
\end{minipage} \ \hspace{.5cm} \
\begin{minipage}[t]{1.2in}
{\small $I=\ccor{\ds{m,1}\ds{n,0}\ds{\ell,0}}{1}$}\vspace{1mm}
\begin{tabular}{|r|r|r|r|}\hline
$m$ & $  n $&  $ \ell $&  $I  $\\ \hline \hline  
1 & 2 & 3 & $  -{7\over {12}}$\\ \hline
1 & 2 & 5 & $  -{7\over {96}}$\\ \hline
1 & 3 & 4 &$   {3\over 8}$\\ \hline
1 & 3 & 6 &$   {{47}\over {54}}$\\ \hline
1 & 4 & 5 &$   {{281}\over {192}}$\\ \hline
1 & 5 & 6 &$   {{3091}\over {3456}}$\\ \hline
2 & 2 & 2 &$   -{7\over {24}}$\\ \hline
2 & 2 & 4 &$   -{1\over 4}$\\ \hline
2 & 2 & 6 & $  {{185}\over {864}}$\\ \hline
2 & 3 & 3 & $  {1\over 4}$\\ \hline
2 & 3 & 5 & $  {{41}\over {24}}$\\ \hline
2 & 4 & 4 & $  {{767}\over {384}}$\\ \hline
2 & 4 & 6 & $  {{4631}\over {3456}}$\\ \hline
2 & 5 & 5 & $  {{727}\over {384}}$\\ \hline
2 & 6 & 6 & $  {{4799}\over {7776}}$\\ \hline
3 & 2 & 3 & $  -{1\over 4}$\\ \hline
3 & 2 & 5 & $  {{49}\over {192}}$\\ \hline
3 & 3 & 4 &$   {{139}\over {96}}$\\ \hline
3 & 3 & 6 & $  {{73}\over {72}}$\\ \hline
3 & 4 & 5 &$   {{217}\over {128}}$\\ \hline
3 & 5 & 6 &$   {{611}\over {768}}$\\ \hline
4 & 2 & 2 & $  -{5\over {48}}$\\ \hline
4 & 2 & 4 &$   {{77}\over {384}}$\\ \hline
4 & 2 & 6 &$   {{215}\over {1152}}$\\ \hline
4 & 3 & 3 &$   {{47}\over {48}}$\\ \hline
4 & 3 & 5 &$   {{37}\over {32}}$\\ \hline
4 & 4 & 4 & $  {{527}\over {384}}$\\ \hline
4 & 4 & 6 &$   {{995}\over {1536}}$\\ \hline
4 & 5 & 5 & $  {{709}\over {768}}$\\ \hline
4 & 6 & 6 & $  {{7903}\over {31104}}$\\ \hline
5 & 2 & 3 & $  {{13}\over {144}}$\\ \hline
5 & 2 & 5 & $  {{187}\over {1152}}$\\ \hline
5 & 3 & 4 & $  {{23}\over {32}}$\\ \hline
5 & 3 & 6 & $  {{443}\over {1296}}$\\ \hline
5 & 4 & 5 & $  {{661}\over {1152}}$\\ \hline
5 & 5 & 6 &$   {{3125}\over {13824}}$\\ \hline
6 & 2 & 2 & $  {5\over {864}}$\\ \hline
6 & 2 & 4 &$   {{83}\over {864}}$\\ \hline
6 & 2 & 6 &$   {{1501}\over {31104}}$\\ \hline
6 & 3 & 3 & $  {{151}\over {432}}$\\ \hline
6 & 3 & 5 &$   {5\over {18}}$\\ \hline
6 & 4 & 4 & $  {{1513}\over {4608}}$\\ \hline
6 & 4 & 6 & $  {{15899}\over {124416}}$\\ \hline
6 & 5 & 5 &$   {{283}\over {1536}}$\\ \hline
6 & 6 & 6 & $  {{1393}\over {31104}}$\\ \hline
\end{tabular}
\end{minipage} \ \hspace{.5cm} \
\begin{minipage}[t]{1.2in}
{\small $I= \ccor{\ds{m,1}\ds{n,1}\ds{\ell,0}}{1}$}\vspace{1mm}
\begin{tabular}{|r|r|r|r|}\hline
$m$&$  n$&$  \ell$&$  I$\\ \hline \hline
1 & 1 & 3 &  $ {1\over {12}}$\\ \hline
1 & 1 & 5 &  $ -{2\over 3}$\\ \hline
1 & 2 & 2 &  $ {1\over 4}$\\ \hline
1 & 2 & 4 &  $ -{{27}\over {32}}$\\ \hline
1 & 2 & 6 &   $-{{631}\over {864}}$\\ \hline
1 & 3 & 3 &   $-{7\over {12}}$\\ \hline
1 & 3 & 5 &   $-{{11}\over {12}}$\\ \hline
1 & 4 & 2 &   $-{5\over {96}}$\\ \hline
1 & 4 & 4 &   $-{{71}\over {96}}$\\ \hline
1 & 4 & 6 &   $-{{1339}\over {3456}}$\\ \hline
1 & 5 & 3 &   $-{{55}\over {144}}$\\ \hline
1 & 5 & 5 &   $-{{197}\over {576}}$\\ \hline
1 & 6 & 2 &   $-{7\over {144}}$\\ \hline
1 & 6 & 4 &   $-{{227}\over {1152}}$\\ \hline
1 & 6 & 6 &   $-{{841}\over {10368}}$\\ \hline
2 & 2 & 3 &   $-{{11}\over {12}}$\\ \hline
2 & 2 & 5 &   $-{3\over 2}$\\ \hline
2 & 3 & 2 &   $-{1\over {24}}$\\ \hline
2 & 3 & 4 &   $-{{253}\over {192}}$\\ \hline
2 & 3 & 6 &   $-{{1255}\over {1728}}$\\ \hline
2 & 4 & 3 &   $-{{43}\over {48}}$\\ \hline
2 & 4 & 5 &   $-{{107}\over {128}}$\\ \hline
2 & 5 & 2 &   $-{1\over 9}$\\ \hline
2 & 5 & 4 &   $-{{599}\over {1152}}$\\ \hline
2 & 5 & 6 &   $-{{2315}\over {10368}}$\\ \hline
2 & 6 & 3 &   $-{{109}\over {432}}$\\ \hline
2 & 6 & 5 &   $-{{157}\over {864}}$\\ \hline
3 & 3 & 3 &   $-{{47}\over {48}}$\\ \hline
3 & 3 & 5 &   $-{{89}\over {96}}$\\ \hline
3 & 4 & 2 &   $-{{31}\over {192}}$\\ \hline
3 & 4 & 4 &   $-{{289}\over {384}}$\\ \hline
3 & 4 & 6 &   $-{{2251}\over {6912}}$\\ \hline
3 & 5 & 3 &   $-{{19}\over {48}}$\\ \hline
3 & 5 & 5 &   $-{{37}\over {128}}$\\ \hline
3 & 6 & 2 &   $-{{47}\over {864}}$\\ \hline
3 & 6 & 4 &   $-{{379}\over {2304}}$\\ \hline
3 & 6 & 6 &   $-{{3803}\over {62208}}$\\ \hline
4 & 4 & 3 &   $-{{33}\over {64}}$\\ \hline
4 & 4 & 5 &   $-{{73}\over {192}}$\\ \hline
4 & 5 & 2 &   $-{{29}\over {384}}$\\ \hline
4 & 5 & 4 &   $-{{181}\over {768}}$\\ \hline
4 & 5 & 6 &   $-{{3679}\over {41472}}$\\ \hline
4 & 6 & 3 &   $-{{197}\over {1728}}$\\ \hline
4 & 6 & 5 &   $-{{1001}\over {13824}}$\\ \hline
5 & 5 & 3 &   $-{{215}\over {1728}}$\\ \hline
5 & 5 & 5 &   $-{{91}\over {1152}}$\\ \hline
5 & 6 & 2 &   $-{{23}\over {1296}}$\\ \hline
5 & 6 & 4 &   $-{{1847}\over {41472}}$\\ \hline
5 & 6 & 6 &   $-{{631}\over {41472}}$\\ \hline
6 & 6 & 3 &   $-{{335}\over {15552}}$\\ \hline
6 & 6 & 5 &   $-{{97}\over {7776}}$\\ \hline
\end{tabular}
\end{minipage}}

%%%%%%%%%%%%%%%
%   Genus-2
%%%%%%%%%%%%%%%
\newpage
\subsection{Genus-Two Descendants} \label{app:list g=2}

\vspace{2mm}
\noindent
\underline{{\sc 1-Point Descendants}}
\begin{center}
{\tiny
\begin{minipage}[t]{2in}
\begin{tabular}{|r|c|}\hline
$n$  &$ \ccor{\ds{n,0}}{2}$\\ \hline \hline
3  &$  -{1\over {240}}$\\ \hline
5  &$  {1\over {576}}$\\ \hline
7  &$  -{{61}\over {23040}}$\\ \hline
9   &$ -{{977}\over {622080}}$\\ \hline
11&$   -{{551}\over {2211840}}$\\ \hline
13 &$  -{{15913}\over {829440000}}$\\ \hline
15 &$  -{{26407}\over {29859840000}}$\\ \hline
17&$   -{{39911}\over {1463132160000}}$\\ \hline
19&$   -{{158371}\over {262193283072000}}$\\ \hline
21&$   -{{9635659}\over {955694516797440000}}$\\ \hline
23&$   -{{4189489}\over {31856483893248000000}}$\\ \hline
25&$   -{{8347673}\over {6057282865987584000000}}$\\ \hline
\end{tabular}
\end{minipage} \ \hspace{5mm} \
\begin{minipage}[t]{3in}
\begin{tabular}{|r|c|}\hline
$n $  &$  \ccor{\ds{n,1}}{2}$\\ \hline \hline
2 &$  {7\over {5760}}$\\ \hline
4 &$  {1\over {1920}}$\\ \hline
6 &$  {{13}\over {7680}}$\\ \hline
8 &$  {{23}\over {41472}}$\\ \hline
10&$   {{77}\over {1105920}}$\\ \hline
12 &$  {{43}\over {9216000}}$\\ \hline
14&$   {{583}\over {2985984000}}$\\ \hline
16 &$  {{13}\over {2322432000}}$\\ \hline
18 &$  {{73}\over {624269721600}}$\\ \hline
20&$   {{1411}\over {758487711744000}}$\\ \hline
22&$   {{589}\over {25282923724800000}}$\\ \hline
24&$   {{103}\over {437033395814400000}}$\\ \hline
\end{tabular}
\end{minipage}}
\end{center}

%%%%%%%%%%%
%   Two Point Functions
%%%%%%%%%%%
\vspace{2mm}
\noindent
\underline{{\sc 2-Point Descendants}}

\noindent
\begin{center}
{\tiny
\begin{minipage}[t]{1.5in}
\begin{tabular}{|r|r|r|}\hline
$m$ &$  n $&$   \ccor{\ds{m,0}\ds{n,0}}{2}$\\ \hline\hline   
2 & 2 &$  -{5\over {288}}$\\ \hline
2 & 4 &$  {{41}\over {2880}}$\\ \hline
2 & 6 &$  -{{59}\over {2560}}$\\ \hline
2 & 8 &$  -{{83}\over {10368}}$\\ \hline
2 & 10 &$  -{{77}\over {2211840}}$\\ \hline
3 & 3 &$  {{29}\over {1440}}$\\ \hline
3 & 5 &$  -{{271}\over {7680}}$\\ \hline
3 & 7 &$  {{211}\over {62208}}$\\ \hline
3 & 9 &$  {{75019}\over {6635520}}$\\ \hline
4 & 4 &$  -{{49}\over {1440}}$\\ \hline
4 & 6 &$  {{1589}\over {69120}}$\\ \hline
4 & 8 &$  {{28027}\over {829440}}$\\ \hline
4 & 10 &$  {{4227469}\over {497664000}}$\\ \hline
5 & 5 &$  {{161}\over {3840}}$\\ \hline
5 & 7 &$  {{21173}\over {311040}}$\\ \hline
5 & 9 &$  {{139061}\over {6635520}}$\\ \hline
6 & 6 &$  {{33377}\over {414720}}$\\ \hline
6 & 8 &$  {{54833}\over {1658880}}$\\ \hline
6 & 10 &$  {{19399463}\over {3583180800}}$\\ \hline
7 & 7 &$  {{9899}\over {248832}}$\\ \hline
7 & 9 &$  {{5954869}\over {716636160}}$\\ \hline
8 & 8 &$  {{185363}\over {19906560}}$\\ \hline
8 & 10 &$  {{7319723}\over {5971968000}}$\\ \hline
9 & 9 &$  {{375793}\over {268738560}}$\\ \hline
10 & 10 &$  {{60039767}\over {429981696000}}$\\ \hline
\end{tabular}
\end{minipage} \ \hspace{5mm} \
\begin{minipage}[t]{1.5in}
\begin{tabular}{|r|r|r|}\hline
$m$&$   n$&$    \ccor{\ds{m,1}\ds{n,0}}{2}$\\ \hline \hline
1 & 2 &$  {7\over {1920}}$\\ \hline 
1 & 4 &$  {1\over {384}}$\\ \hline 
1 & 6 &$  -{9\over {2560}}$\\ \hline 
1 & 8 & $ -{{901}\over {69120}}$\\ \hline 
1 & 10 &$  -{{4499}\over {1105920}}$\\ \hline 
2 & 3 & $ {1\over {192}}$\\ \hline 
2 & 5 &$  -{{47}\over {7680}}$\\ \hline 
2 & 7 &$  -{{29}\over {864}}$\\ \hline 
2 & 9 &$  -{{139307}\over {9953280}}$\\ \hline 
3 & 2 &$  {1\over {192}}$\\ \hline 
3 & 4 &$  {{13}\over {4608}}$\\ \hline 
3 & 6 &$  -{{4931}\over {103680}}$\\ \hline 
3 & 8 &$  -{{30409}\over {1105920}}$\\ \hline 
3 & 10 &$  -{{672173}\over {124416000}}$\\ \hline 
4 & 3 &$  {{89}\over {7680}}$\\ \hline 
4 & 5 &$  -{{1117}\over {23040}}$\\ \hline 
4 & 7 &$  -{{34397}\over {829440}}$\\ \hline 
4 & 9 &$  -{{69811}\over {6635520}}$\\ \hline 
5 & 2 &$  {{91}\over {7680}}$\\ \hline 
5 & 4 &$  -{{49}\over {1920}}$\\ \hline 
5 & 6 &$  -{{423}\over {10240}}$\\ \hline 
5 & 8 &$  -{{142513}\over {9953280}}$\\ \hline 
5 & 10 &$  -{{3204979}\over {1492992000}}$\\ \hline 
6 & 3 &$  -{{811}\over {103680}}$\\ \hline 
6 & 5 & $ -{{5107}\over {165888}}$\\ \hline 
6 & 7 &$  -{{353}\over {23040}}$\\ \hline 
6 & 9 &$  -{{35275}\over {11943936}}$\\ \hline 
7 & 2 &$  {{23}\over {11520}}$\\ \hline 
7 & 4 &$  -{{1303}\over {92160}}$\\ \hline 
7 & 6 &$  -{{1613}\over {138240}}$\\ \hline 
7 & 8 &$  -{{40423}\over {13271040}}$\\ \hline 
7 & 10 &$  -{{126889}\over {331776000}}$\\ \hline 
8 & 3 &$  -{{37}\over {8192}}$\\ \hline 
8 & 5 &$  -{{1525}\over {221184}}$\\ \hline 
8 & 7 &$  -{{302807}\over {119439360}}$\\ \hline 
8 & 9 &$  -{{292267}\over {716636160}}$\\ \hline 
9 & 2 &$  -{{121}\over {1105920}}$\\ \hline 
9 & 4 &$  -{{43769}\over {16588800}}$\\ \hline 
9 & 6 &$  -{{186277}\over {119439360}}$\\ \hline 
9 & 8 &$  -{{33649}\over {99532800}}$\\ \hline 
9 & 10 &$  -{{1071761}\over {28665446400}}$\\ \hline 
10 & 3 &$  -{{10163}\over {13824000}}$\\ \hline 
10 & 5 &$  -{{253273}\over {331776000}}$\\ \hline 
10 & 7 &$  -{{86129}\over {373248000}}$\\ \hline 
10 & 9 &$  -{{1557227}\over {47775744000}}$\\ \hline 
\end{tabular}
\end{minipage} \ \hspace{5mm} \
\begin{minipage}[t]{1.5in}
\begin{tabular}{|r|r|r|}\hline
$m$&$   n$&$    \ccor{\ds{m,1}\ds{n,1}}{2}$ \\ \hline \hline
1 & 5 & $ {{91}\over {11520}}$ \\ \hline
1 & 7 &$  {{23}\over {3840}}$ \\ \hline
1 & 9 &$  {{121}\over {92160}}$ \\ \hline
2 & 2 & $ {1\over {576}}$ \\ \hline
2 & 4 &$  {{71}\over {3840}}$ \\ \hline
2 & 6 &$  {{67}\over {3840}}$ \\ \hline
2 & 8 & $ {{809}\over {165888}}$ \\ \hline
2 & 10 &$  {{1711}\over {2764800}}$ \\ \hline
3 & 3 & $ {{25}\over {1152}}$ \\ \hline
3 & 5 &$  {{1001}\over {34560}}$ \\ \hline
3 & 7 &$  {{319}\over {30720}}$ \\ \hline
3 & 9 &$  {{209}\over {129600}}$ \\ \hline
4 & 4 &$  {{421}\over {11520}}$ \\ \hline
4 & 6 &$  {{149}\over {8640}}$ \\ \hline
4 & 8 &$  {{3697}\over {1105920}}$ \\ \hline
4 & 10 &$  {{56371}\over {165888000}}$ \\ \hline
5 & 5 &$  {{451}\over {23040}}$ \\ \hline
5 & 7 &$  {{4061}\over {829440}}$ \\ \hline
5 & 9 &$  {{60523}\over {99532800}}$\\ \hline
6 & 6 &$  {{791}\over {138240}}$ \\ \hline
6 & 8 & $ {{17651}\over {19906560}}$ \\ \hline
6 & 10 &$  {{38407}\over {497664000}}$ \\ \hline
7 & 7 &$  {{1627}\over {1658880}}$ \\ \hline
7 & 9 &$  {{3467}\over {33177600}}$ \\ \hline
8 & 8 & $ {{7037}\over {59719680}}$ \\ \hline
8 & 10 &$  {{1823}\over {199065600}}$ \\ \hline
9 & 9 & $ {{2377}\over {238878720}}$ \\ \hline
10 & 10 &$  {{15589}\over {23887872000}}$ \\ \hline
\end{tabular}
\end{minipage}}
\end{center}

%%%%%%%%%%
%  Three Point Functions
%%%%%%%%%%

\newpage
\vspace{2mm}
\noindent
\underline{{\sc 3-Point Descendants}}
\vspace{2mm}

\noindent
\begin{center}
\noindent
{\tiny
\begin{minipage}[t]{1.2in}
{\small $I=\ccor{\ds{m,0}\ds{n,0}\ds{\ell,0}}{2}$}\vspace{1mm}
\begin{tabular}{|r|r|r|r|}\hline
$m$ & $ n$ & $ \ell$ &$ I$\\ \hline\hline
2 & 2 & 3 & $  {{11}\over {96}}$\\ \hline
2 & 2 & 5 & $  -{{239}\over {1280}}$\\ \hline
2 & 3 & 4 & $  -{{1411}\over {5760}}$\\ \hline
2 & 3 & 6 & $  {{4211}\over {25920}}$\\ \hline
2 & 4 & 5 & $  {{7147}\over {23040}}$\\ \hline
2 & 5 & 6 & $  {{162199}\over {414720}}$\\ \hline
2 & 6 & 5 & $  {{162199}\over {414720}}$\\ \hline
3 & 3 & 3 & $  -{1\over 3}$\\ \hline
3 & 3 & 5 & $  {4\over 9}$\\ \hline
3 & 4 & 4 &$   {{569}\over {1152}}$\\ \hline
3 & 4 & 6 &$   {{1247}\over {5760}}$\\ \hline
3 & 5 & 5 & $  {{2017}\over {11520}}$\\ \hline
3 & 6 & 6 & $  -{{43471}\over {69120}}$\\ \hline
4 & 4 & 5 & $  {{971}\over {15360}}$\\ \hline
4 & 5 & 6 & $  -{{1040729}\over {829440}}$\\ \hline
5 & 5 & 5 & $  -{{1709}\over {960}}$\\ \hline
5 & 6 & 6 & $  -{{1547117}\over {995328}}$\\ \hline
\end{tabular}
\end{minipage} \ \hspace{.5cm} \
\begin{minipage}[t]{1.2in}
{\small $I=\ccor{\ds{m,1}\ds{n,1}\ds{\ell,1}}{2}$}\vspace{1mm}
\begin{tabular}{|r|r|r|r|}\hline
$m$ & $n$&  $\ell $ &  $I$\\ \hline \hline
1 & 1 & 4 &  $  {{61}\over {1920}}$\\ \hline
1 & 1 & 6 &   $ {{113}\over {1920}}$\\ \hline
1 & 2 & 3 &  $  {{35}\over {576}}$\\ \hline
1 & 2 & 5 &  $  {{413}\over {2880}}$\\ \hline
1 & 3 & 4 &  $  {{821}\over {3840}}$\\ \hline
1 & 3 & 6 &  $  {{13}\over {90}}$\\ \hline
1 & 4 & 5 &  $  {{589}\over {2880}}$\\ \hline
1 & 5 & 6 &  $  {{18247}\over {207360}}$\\ \hline
2 & 2 & 2 &  $  {{19}\over {192}}$\\ \hline
2 & 2 & 4 &  $  {{1217}\over {3840}}$\\ \hline
2 & 2 & 6 &  $  {{22547}\over {103680}}$\\ \hline
2 & 3 & 3 &  $  {{107}\over {288}}$\\ \hline
2 & 3 & 5 &  $  {{12473}\over {34560}}$\\ \hline
2 & 4 & 4 &  $  {{5203}\over {11520}}$\\ \hline
2 & 4 & 6 &  $  {{27499}\over {138240}}$\\ \hline
2 & 5 & 5 &  $  {{1561}\over {6912}}$\\ \hline
2 & 6 & 6 &  $  {{9187}\over {138240}}$\\ \hline
3 & 3 & 4 & $   {{12193}\over {23040}}$\\ \hline
3 & 3 & 6 & $   {{48271}\over {207360}}$\\ \hline
3 & 4 & 5 & $   {{22799}\over {69120}}$\\ \hline
3 & 5 & 6 & $   {{22859}\over {207360}}$\\ \hline
4 & 4 & 4 & $   {{4217}\over {10240}}$\\ \hline
4 & 4 & 6 &  $  {{38239}\over {276480}}$\\ \hline
4 & 5 & 5 &  $  {{130099}\over {829440}}$\\ \hline
4 & 6 & 6 & $   {{38257}\over {995328}}$\\ \hline
5 & 5 & 6 &   $   {{36073}\over {829440}}$\\ \hline
6 & 6 & 6 &  $  {{15431}\over {1658880}}$\\ \hline
\end{tabular}
\end{minipage} \ \hspace{.5cm} \
\begin{minipage}[t]{1.2in}
{\small $I=\ccor{\ds{m,1}\ds{n,0}\ds{\ell,0}}{2}$}\vspace{1mm}
\begin{tabular}{|r|r|r|r|}\hline
$m$ & $  n $&  $ \ell $&  $I  $\\ \hline \hline 
1 & 2 & 3 & $   {1\over {32}}$\\ \hline
1 & 2 & 5 & $   -{{317}\over {5760}}$\\ \hline
1 & 3 & 4 & $   -{{361}\over {5760}}$\\ \hline
1 & 3 & 6 & $   -{{6703}\over {51840}}$\\ \hline
1 & 4 & 5 & $   -{{2021}\over {23040}}$\\ \hline
1 & 5 & 6 & $   {{48073}\over {103680}}$\\ \hline
2 & 2 & 2 & $   {{23}\over {576}}$\\ \hline
2 & 2 & 4 & $   -{{47}\over {1280}}$\\ \hline
2 & 2 & 6 &  $  -{{4259}\over {17280}}$\\ \hline
2 & 3 & 3 & $   -{7\over {160}}$\\ \hline
2 & 3 & 5 & $   -{{35}\over {144}}$\\ \hline
2 & 4 & 4 & $   -{{2381}\over {11520}}$\\ \hline
2 & 4 & 6 & $   {{7969}\over {15360}}$\\ \hline
2 & 5 & 5 & $   {{18077}\over {23040}}$\\ \hline
2 & 6 & 6 & $   {{133319}\over {138240}}$\\ \hline
3 & 2 & 3 & $   {{11}\over {576}}$\\ \hline
3 & 2 & 5 & $   -{{3521}\over {11520}}$\\ \hline
3 & 3 & 4 & $   -{{107}\over {360}}$\\ \hline
3 & 3 & 6 & $   {{4081}\over {17280}}$\\ \hline
3 & 4 & 5 & $   {{27301}\over {46080}}$\\ \hline
3 & 5 & 6 & $   {{48989}\over {41472}}$\\ \hline
3 & 6 & 5 & $   {{48989}\over {41472}}$\\ \hline
4 & 2 & 2 & $   {{53}\over {768}}$\\ \hline
4 & 2 & 4 & $   -{{1399}\over {5760}}$\\ \hline
4 & 2 & 6 &  $  -{{20933}\over {138240}}$\\ \hline
4 & 3 & 3 & $   -{{383}\over {1440}}$\\ \hline
4 & 3 & 5 & $   {{331}\over {1536}}$\\ \hline
4 & 4 & 4 & $   {{6605}\over {18432}}$\\ \hline
4 & 4 & 6 & $   {{147143}\over {165888}}$\\ \hline
4 & 5 & 5 & $   {{37829}\over {30720}}$\\ \hline
4 & 6 & 6 & $   {{2502289}\over {2985984}}$\\ \hline
5 & 2 & 3 & $   -{{413}\over {3456}}$\\ \hline
5 & 2 & 5 & $   -{{9617}\over {69120}}$\\ \hline
5 & 3 & 4 & $   {{179}\over {2560}}$\\ \hline
5 & 3 & 6 & $   {{1805}\over {4608}}$\\ \hline
5 & 4 & 3 & $   {{179}\over {2560}}$\\ \hline
5 & 4 & 5 & $   {{101917}\over {138240}}$\\ \hline
5 & 5 & 6 & $   {{596827}\over {829440}}$\\ \hline
6 & 2 & 2 & $   -{{373}\over {34560}}$\\ \hline
6 & 2 & 4 & $   -{{7121}\over {82944}}$\\ \hline
6 & 2 & 6 & $   -{{101}\over {23040}}$\\ \hline
6 & 3 & 3 & $   -{{143}\over {10368}}$\\ \hline
6 & 3 & 5 & $   {{9827}\over {34560}}$\\ \hline
6 & 4 & 4 & $   {{320921}\over {829440}}$\\ \hline
6 & 4 & 6 & $   {{5923543}\over {14929920}}$\\ \hline
6 & 5 & 5 & $   {{28171}\over {51840}}$\\ \hline
6 & 6 & 6 & $   {{4004861}\over {14929920}}$\\ \hline
\end{tabular}
\end{minipage} \ \hspace{.5cm} \
\begin{minipage}[t]{1.2in}
{\small $I= \ccor{\ds{m,1}\ds{n,1}\ds{\ell,0}}{2}$}\vspace{1mm}
\begin{tabular}{|r|r|r|r|}\hline
$m$&$  n$&$  \ell$&$  I$\\ \hline \hline
1 & 1 & 5 & $  {{17}\over {720}}$\\ \hline
1 & 2 & 2 &$   {1\over {192}}$\\ \hline
1 & 2 & 4 &$   {{61}\over {1152}}$\\ \hline
1 & 2 & 6 &$   -{{397}\over {2304}}$\\ \hline
1 & 3 & 3 & $  {{17}\over {192}}$\\ \hline
1 & 3 & 5 & $  -{{551}\over {2880}}$\\ \hline
1 & 4 & 2 &$   {{151}\over {1920}}$\\ \hline
1 & 4 & 4 & $  -{{71}\over {720}}$\\ \hline
1 & 4 & 6 & $  -{{25687}\over {69120}}$\\ \hline
1 & 5 & 3 &$   {7\over {4320}}$\\ \hline
1 & 5 & 5 &$   -{{4201}\over {13824}}$\\ \hline
1 & 6 & 2 & $  {{167}\over {3840}}$\\ \hline
1 & 6 & 4 & $  -{{533}\over {3456}}$\\ \hline
1 & 6 & 6 & $  -{{26251}\over {138240}}$\\ \hline
2 & 2 & 3 &$   {{11}\over {96}}$\\ \hline
2 & 2 & 5 &$   -{{3283}\over {11520}}$\\ \hline
2 & 3 & 2 & $  {{17}\over {128}}$\\ \hline
2 & 3 & 4 &$   -{{25}\over {144}}$\\ \hline
2 & 3 & 6 &$   -{{26813}\over {41472}}$\\ \hline
2 & 4 & 3 &$   -{1\over {180}}$\\ \hline
2 & 4 & 5 &$   -{{1019}\over {1536}}$\\ \hline
2 & 5 & 2 &$   {{1267}\over {11520}}$\\ \hline
2 & 5 & 4 &$   -{{1733}\over {4608}}$\\ \hline
2 & 5 & 6 &$   -{{199573}\over {414720}}$\\ \hline
2 & 6 & 3 &$   -{{6767}\over {51840}}$\\ \hline
2 & 6 & 5 &$   -{{3313}\over {9216}}$\\ \hline
3 & 3 & 3 &$   -{1\over {144}}$\\ \hline
3 & 3 & 5 &$   -{{8957}\over {11520}}$\\ \hline
3 & 4 & 2 & $  {{913}\over {5760}}$\\ \hline
3 & 4 & 4 &$   -{{3203}\over {5760}}$\\ \hline
3 & 4 & 6 & $  -{{29155}\over {41472}}$\\ \hline
3 & 5 & 3 & $    -{{2501}\over {11520}}$\\ \hline
3 & 5 & 5 & $  -{{82409}\over {138240}}$\\ \hline
3 & 6 & 2 & $  {{413}\over {13824}}$\\ \hline
3 & 6 & 4 & $  -{{44633}\over {138240}}$\\ \hline
3 & 6 & 6 & $  -{{646351}\over {2488320}}$\\ \hline
4 & 4 & 3 &$   -{{11}\over {40}}$\\ \hline
4 & 4 & 5 &$   -{{11419}\over {15360}}$\\ \hline
4 & 5 & 2 & $  {{973}\over {23040}}$\\ \hline
4 & 5 & 4 & $  -{{126253}\over {276480}}$\\ \hline
4 & 5 & 6 & $  -{{458771}\over {1244160}}$\\ \hline
4 & 6 & 3 & $  -{{6419}\over {34560}}$\\ \hline
4 & 6 & 5 & $  -{{58183}\over {207360}}$\\ \hline
5 & 5 & 3 & $  -{{27}\over {128}}$\\ \hline
5 & 5 & 5 &$   -{{1223}\over {3840}}$\\ \hline
5 & 6 & 2 &$   -{{559}\over {82944}}$\\ \hline
5 & 6 & 4 &$   -{{2747}\over {15552}}$\\ \hline
5 & 6 & 6 & $  -{{60367}\over {552960}}$\\ \hline
6 & 6 & 3 &$   -{{47233}\over {622080}}$\\ \hline
6 & 6 & 5 & $  -{{1255591}\over {14929920}}$\\ \hline
\end{tabular}
\end{minipage}}
\end{center}

%%%%%%%%%%%%%%%%%#########################################
%  References
%%%%%%%%%%%%%%%%%#########################################
\newpage

\nc{\bi}{\bibitem}

\end{document}